\documentclass[aps,prb,twocolumn,longbibliography,superscriptaddress]{revtex4-1}
\usepackage{epsfig}
\usepackage{epstopdf}
\usepackage{amsmath}
\usepackage{amsfonts}
\usepackage{amssymb}
\usepackage{hyperref}
\usepackage{bm}
\usepackage{bbm}
\usepackage{makecell}
\usepackage{rotating}
\usepackage{hyperref}
\usepackage{dsfont}
\usepackage{mathdots}
\usepackage{latexsym}
\usepackage[normalem]{ulem}
\usepackage{mathtools}

\usepackage{graphicx}% Include figure files
\usepackage{dcolumn}% Align table columns on decimal point
\usepackage{bm}% bold math
\usepackage{color}

\usepackage{tikz,xcolor,hyperref}

\definecolor{lime}{HTML}{A6CE39}
\DeclareRobustCommand{\orcidicon}{%
	\begin{tikzpicture}
	\draw[lime, fill=lime] (0,0)
	circle [radius=0.16]
	node[white] {{\fontfamily{qag}\selectfont \tiny ID}};
	\draw[white, fill=white] (-0.0625,0.095)
	circle [radius=0.007];
	\end{tikzpicture}
	\hspace{-2mm}
}

\foreach \x in {A, ..., Z}{%
	\expandafter\xdef\csname orcid\x\endcsname{\noexpand\href{https://orcid.org/\csname orcidauthor\x\endcsname}{\noexpand\orcidicon}}
}

\begin{document}

\title{Dual topology and edge-reconstruction in $\alpha$-Sn}
%\title{Unexpected edge-states due to approximate winding numbers in layered $\alpha$-Sn}
%\title{Topological properties of the multi-layer $\alpha$-Sn}

\author{Jan Skolimowski\orcidG}
\affiliation{International Research Centre Magtop, Institute of Physics, Polish Academy of Sciences,
Aleja Lotnik\'ow 32/46, PL-02668 Warsaw, Poland}

\author{Nguyen Minh Nguyen\orcidF}
\affiliation{School of Science and Engineering, The Chinese University of Hong Kong, Shenzhen, Shenzhen 518172, China}
\affiliation{International Research Centre Magtop, Institute of Physics, Polish Academy of Sciences,
Aleja Lotnik\'ow 32/46, PL-02668 Warsaw, Poland}

\author{Alicja Kawala}
\affiliation{Institute of Theoretical Physics, Jagiellonian University, ulica S. \L{}ojasiewicza 11, PL-30348 Krak\'ow, Poland}
\affiliation{Jagiellonian University, Doctoral School of Exact and Natural Sciences}

\author{Giuseppe Cuono\orcidD}
\affiliation{Consiglio Nazionale delle Ricerche (CNR-SPIN), Unit\'a di Ricerca presso Terzi c/o Universit\'a “G. D’Annunzio”, 66100 Chieti, Italy}

\author{Carmine Autieri\orcidA}
%\email{autieri@magtop.ifpan.edu.pl}
\affiliation{International Research Centre Magtop, Institute of Physics, Polish Academy of Sciences,
Aleja Lotnik\'ow 32/46, PL-02668 Warsaw, Poland}
\affiliation{SPIN-CNR, UOS Salerno, IT-84084 Fisciano (SA), Italy}

\author{Wojciech Brzezicki\orcidC}
\affiliation{International Research Centre Magtop, Institute of Physics, Polish Academy of Sciences,
Aleja Lotnik\'ow 32/46, PL-02668 Warsaw, Poland}

%\author{Tomasz Dietl\orcidB}
%\affiliation{International Research Centre Magtop, Institute of Physics, Polish Academy of Sciences, Aleja Lotnik\'ow 32/46, PL-02668 Warsaw, Poland}
%\affiliation{WPI-Advanced Institute for Materials Research, Tohoku University, Sendai 980-8577, Japan}

\date{\today}
\begin{abstract}
We formulate the tight-binding model for cubic $\alpha$-Sn based on the DFT calculations. In the model, we incorporate a variable bond angle, which allows us to simulate the effect of the in-plane strain. 
In the bulk, we demonstrate the presence of the $\mathbb{Z}_2$ topological invariant and a non-zero mirror Chern number, making $\alpha$-Sn one of the rare cases where dual topology can be observed.
We calculate the topological phase diagram of multi-layer $\alpha$-Sn as a function of strain and number of layers. We find that a non-trivial quantum spin Hall state appears only for compressive strain above five layers of thickness. Quite surprisingly, both in the trivial and non-trivial phases, we find a plethora of edge-states with energies inside the bulk gap of the system.
Some of these states are localized at the side surfaces of the slab and some of them prefer top/bottom surfaces. We find that they can be adiabatically connected to topological states in the chiral limit, where the model supports multiple one-dimensional winding numbers that take different values in different directions in the Brillouin zone. Finally, around high-symmetry points their presence can also be explained by the mirror Chern number and Wilson-loop invariants.
\end{abstract}


\maketitle

\section{Introduction}

$\alpha$-Sn or gray tin is a semiconducting allotrope of Sn stable at low temperatures. The crystal structure is similar to HgTe (space group No. 216, F$\overline{4}$3m), but since the two atoms of the basis are of the same kind, the $\alpha$-Sn system (space group No. 227, Fd$\overline{3}$m) presents the inversion symmetry in addition to HgTe. It has garnered significant attention because of its non-trivial band topology \cite{Sanchez16,LeDuc21,Ohtsubo13,PhysRevLett.111.157205,PhysRevMaterials.8.044202,PhysRevB.105.075109}. Epitaxial strain can dramatically enhance its stability, allowing thin $\alpha$-Sn films to persist even at elevated temperatures\cite{Thermal_Stability}. Strain not only stabilizes $\alpha$‑Sn but also profoundly reshapes its electronic structure\cite{PhysRevB.95.201101}. Compressive in-plane strain, for example, transforms $\alpha$‑Sn into a three-dimensional (3D) Dirac semimetal, as confirmed by ARPES measurements revealing linear surface Dirac cones and spin-momentum-locked topological states\cite{Polacinski2023} with tunable Dirac points, offering control over novel transport phenomena like negative magnetoresistance\cite{PhysRevB.109.245135}. Alternatively, tensile in-plane strain can induce a topological insulating phase, which has been less studied experimentally. The interplay of strain, thickness, and spin–orbit coupling leads to topological phase transitions also in the thin films of $\alpha$-Sn grown along the (111) orientation\cite{SHI2020126782,PhysRevB.100.245144}. Quantum wells of $\alpha$‑Sn under varying strain exhibit switchable topological phases, highlighting its potential in topological and quantum devices. Capping layers such as AlO$_x$ help preserve the topological surface states by maintaining surface integrity\cite{PhysRevB.98.195445}. Recent work underscores the strain’s central role in tuning the electronic topology, including the quantum spin Hall phase (QSH), enabling room-temperature studies of Dirac fermions and quantum oscillations\cite{Polacinski2023}. A large superconducting diode was observed by interfacing the topological $\alpha$- and the superconducting $\beta$-Sn phase\cite{Anh2024}.

The topological origin of the QSH effect implies the existence 
of helical edge modes, which are protected against 
elastic backscattering from all perturbations obeying time-reversal symmetry. 
Various materials have been predicted to support the QSH insulator phase 
\cite{RevKane, RevZhang, BHZ, Liu08}. 
Experimentally, signatures of edge mode transport have been observed in several of the candidate materials, such as 
HgTe/CdTe quantum wells \cite{QSH_HgTe}, InAs/GaSb bilayers \cite{Du15}, 
WTe$_2$ \cite{Wu18}, bismuthene on SiC \cite{Reis17}
and Moir\'e materials \cite{Kang2024b}.
However, the experimental studies have also led to 
discrepancies with the simple theoretical models, with the protection length of the edge transport reaching only a few tens of nanometers in WTe$_2$ \cite{Wu18}, a few micrometers in InAs/GaSb \cite{Du15} and a few tens of micrometers in HgTe/CdTe quantum wells \cite{PhysRevLett.123.047701}.
There are several possible explanations for this faulty QSH effect, including magnetic impurities \cite{PhysRevLett.106.236402}, phonons \cite{PhysRevLett.108.086602}, dynamic nuclear polarization \cite{PhysRevB.86.035112,PhysRevB.87.165440}, spontaneous time-reversal symmetry breaking \cite{Pikulin14, Paul1, Paul2}, charge puddles \cite{Vayrynen14}, charge dopants \cite{Dietl2022} and interaction effects \cite{Dolcetto16}. A more recent scenario to explain short protection length
is the reconstruction of the charge density at the edges \cite{Wang2017,Amaricci2017,Brze2023,Soni2025}, as 
enhanced edge density of states can trigger local magnetic moment formation. It has been demonstrated that once multiple channels exist \cite{Brze2023}, time-reversal symmetry does not forbid scattering
between different Kramers pairs \cite{Erhardt2025}. 

In this paper, we use first-principles calculations to derive an effective tight-binding model for $\alpha$-Sn including the effects of lattice strain. After establishing its 3D topological properties, we show that this material supports additional surface states on the top and side surfaces of the system,  whose number is proportional to the system's height or width, respectively.
We find that these states can be adiabatically connected to topological edge-states of a chiral-symmetric model with with one-dimensional (1D) winding numbers supporting different numbers of flat bands at different terminations of the system. This limit is achieved when we set the on-site energies of the $s$ and $p$ orbitals to be the same for all sites and we switch off atomic spin-orbit coupling. We show that the flat bands found in this limit smoothly transform  into the additional dispersive edge states found in the spectrum of the full Hamiltonian of $\alpha$-Sn.

\section{Tight-binding parameters extracted from first-principles calculations}

% first-part of the section 2 on DFT
We carried out electronic structure calculations using the VASP package \cite{VASP2}, based on a plane-wave basis set and the projector augmented-wave (PAW) method \cite{VASP}. A plane-wave energy cutoff of 250 eV was adopted for all calculations. The Brillouin zone was sampled using an $8 \times 8 \times 8$ Monkhorst-Pack $k$-point mesh. In the presence of spin-orbit coupling (SOC), 512 $k$-points were used within the irreducible Brillouin zone, whereas 176 $k$-points were utilized in the absence of SOC. For the treatment of exchange-correlation effects, we employed the generalized gradient approximation (GGA) for the exchange-correlation functional.

\begin{figure}[!t]
\centering
\includegraphics[width=4cm,angle=0]{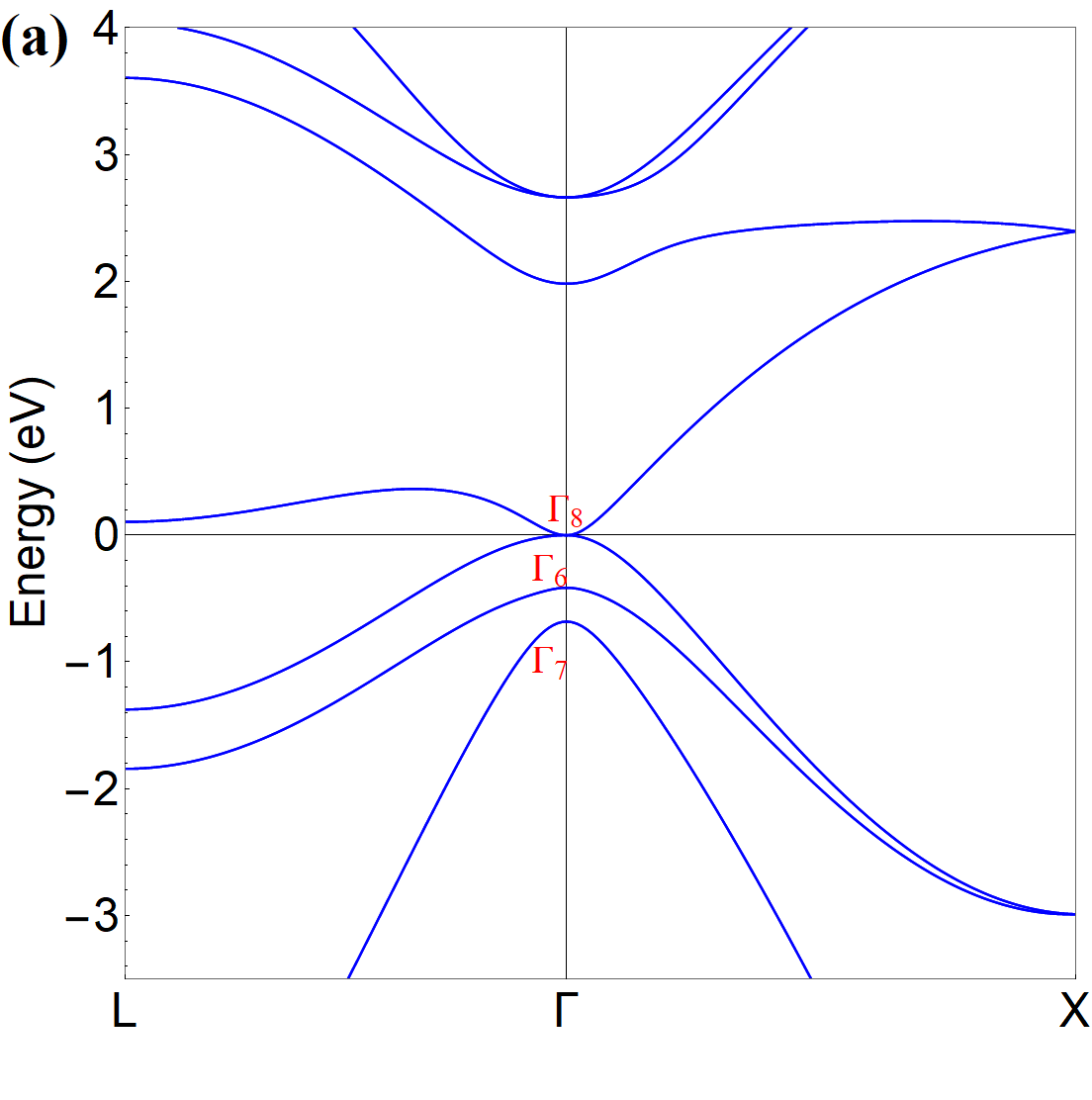}
\includegraphics[width=4cm,angle=0]{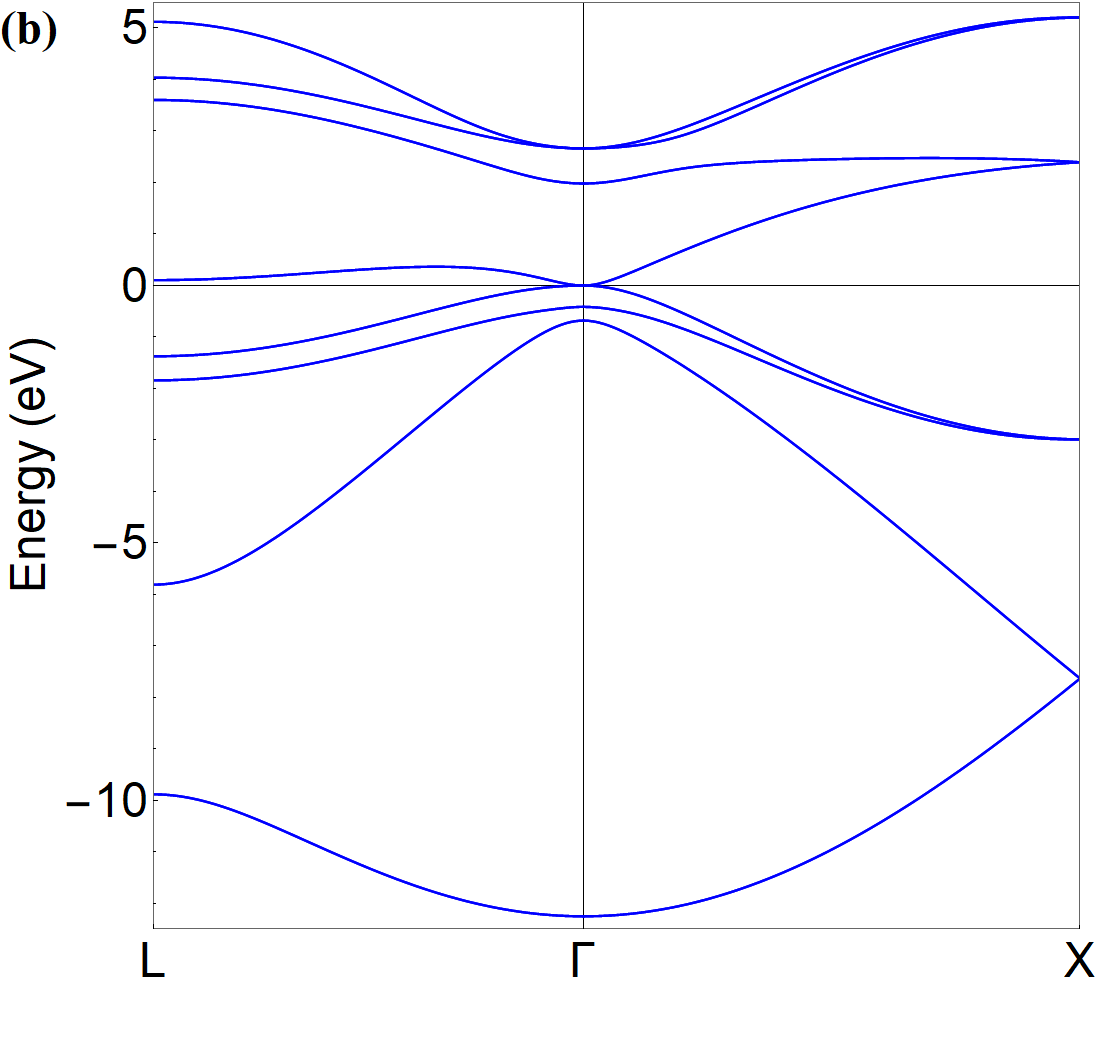}
\caption{Relativistic band structure of $\alpha$-Sn using the tight-binding model reported in Table \ref{table:1}. (a) Band structure in the energy range from -3.5 eV to +4 eV. (b) Band structure in an extended energy range. The X-point is at the coordinates ($\frac{1}{2}$,0,0), while the L-point is at the coordinates ($\frac{1}{2}$,$\frac{1}{2}$,$\frac{1}{2}$). $\Gamma_8$ and $\Gamma_7$ represent the p-states at the $\Gamma$ point while $\Gamma_6$ represents the s-state.
}
\label{Bands_tight_binding}
\end{figure}

%second part of the section 2 on the symmetry of the band structure + error in the gap (DONE)
Because the system is non-magnetic and possesses inversion symmetry, all bands exhibit Kramers degeneracy.
At the $\Gamma$ point, when we include the spin-orbit coupling (SOC), the p-states are separated into the fourfold degenerate $\Gamma_8$ and the twofold degenerate $\Gamma_7$. The $\Gamma_6$ originates from an s-orbital, and the topological nature of the system arises from a band inversion between the $\Gamma_6$ and $\Gamma_8$ energy levels.
{Within the GGA approximation, the calculated band structure does not agree with experiment, as the conduction band minimum at the $L$ point lies below the $\Gamma_8$ state. This GGA error at the $L$ point incorrectly renders the system metallic and without topological invariants~\cite{materialsproject_mp117}, whereas experimentally it is a zero-gap topological semiconductor. In density functional theory, this discrepancy is attributed to the conventional band-gap problem, which in the case of $\alpha$-Sn occurs both with and without spin–orbit coupling. Improved agreement with experimental results at the $L$ point has been achieved using hybrid functionals such as HSE~\cite{Kufner2013structural}, the meta-GGA functional MBJ~\cite{PhysRevB.109.245135}, and relativistic self-consistent GW methods~\cite{Zgid2026}. Once the issue at the L point is corrected, the topological properties are determined by the strength of the band inversion and the ordering of the energy levels at $\Gamma$, which experimentally have the order $\Gamma_8$, $\Gamma_6$, and $\Gamma_7$ as reported in Fig. \ref{Bands_tight_binding}(a). According to the literature, all the functionals mentioned above reproduce the correct ordering, with only minor differences in the magnitude of the band inversion among them. In summary, correcting the conduction-band energy level at the L point is the only modification required to bring the GGA results into qualitative agreement with HSE, MBJ, and self-consistent GW calculations.}

\begin{table}
\begin{center}
    \caption{First-nearest-neighbor tight-binding parameters for cubic diamond $\alpha$-Sn compared with two previous sp$^3$s$^{*}$ tight-binding models\cite{Kufner2013structural,VOGL1983365}. The value of spin-orbit coupling $\lambda$ was taken from the HSE calculations\cite{Kufner2013structural}, considering that the spin-orbit splitting at $\Gamma$ is $\Delta_0$=3/2$\lambda$. 
All values are in eV.\label{table:1}}
    \begin{tabular}{|c|c|c|c|c|}
    \hline
       & This work & sp$^3$s$^{*}$ in Ref. \onlinecite{Kufner2013structural} & sp$^3$s$^{*}$ in Ref. \onlinecite{VOGL1983365} \\
    \hline
    E$_s$                  & -6.335  & -7.2029  &  -5.8700 \\
    \hline
    E$_p$                  &  1.104  &  0.0032  &   1.3300 \\
    \hline
    E$_{s^{*}}$              & -       & 14.1081  &   5.9000 \\
    \hline
    $\lambda$              &  0.454  &  0.454   &      -   \\
    \hline
    V$_{ss\sigma}$         & -1.480  & -1.4802  &  -1.4175 \\
    \hline
    V$_{s^{*}s^{*}\sigma}$ &    -    & 0.4795   &    -     \\
   \hline
    V$_{sp\sigma}$         &  1.458  & 1.9034   &   1.9536 \\
    \hline
    V$_{s^{*}p\sigma}$     &    -    & 2.7564   &   2.5522 \\
     \hline
    V$_{pp\sigma}$         &  2.373  & 3.3555   &   2.3725 \\
     \hline
    V$_{pp\pi}$            & -0.687  & -1.6848  &  -0.6870\\
    \hline
        \end{tabular}
        \end{center}
\end{table}

 %figure1

Now, we move to extract the tight-binding parameters from the GGA calculations by fitting our band structure at the k-points $\Gamma$, X and L, following the same procedure described in the literature\cite{Autieri2021momentum}. 
In order to fit the experimental values, we renormalize three hopping parameters. We use the parameters V$_{pp\sigma}$ and V$_{pp\pi}$ from the literature\cite{VOGL1983365}.
We finally produce our tight-binding by renormalizing V$_{sp\sigma}$ and fitting the experimental information that the conduction band at the L-point is 100 meV above the Fermi level\cite{Polacinski2023}. This procedure decreases the value of V$_{sp\sigma}$; the conduction band at the L point reproduces the experimental results. After shifting the on-site energy to set the zero of the energy at the top of the valence band, the results are reported in the first column of Table \ref{table:1}. The sign and the size of the tight-binding parameters are the same as those of the previous tight-binding model calculated for other cubic zinc-blended semiconductors\cite{Autieri2021momentum,Nguyen2023unprotected}. Using parameters from our tight-binding model, we plot the band structure along the high-symmetry k-path L-$\Gamma$-X in Fig. \ref{Bands_tight_binding} to show the correct behaviour of our results at the L point. Fig. \ref{Bands_tight_binding}(a) shows the low-energy band structure in the range from -3.5 and 4 eV. We can observe how the conduction band at the L point does not cross the Fermi level. In Fig. \ref{Bands_tight_binding}(b), we report the entire energy range where we can observe all the bands obtained within our model. Despite its simplicity, our sp$^3$ model captures the key topological features of the low-energy band structure of $\alpha$-Sn.

% fourth part of section 2 on tight-binding, comparison
{We compare our results on the tight-binding model with theoretical and experimental literature. 
Regarding the tight-binding model, in the theoretical literature, we have \emph{k}$\cdot$\emph{p} models and two different sp$^3$s$^{*}$ models\cite{Kufner2013structural,VOGL1983365} for $\alpha$-Sn, which are reported in Table \ref{table:1}. In this paper, we go beyond producing a sp$^3$ model for $\alpha$-Sn. In the previous sp$^3$s$^{*}$ models, the s-orbital in the model represents the 5s orbitals of Sn, while the s$^{*}$ represents the 6s orbitals; however, this s$^{*}$ orbital can be removed without affecting too much the quality of the tight-binding model and simplifying the model. There are two reasons why we can remove the s$^{*}$ from the model. The first is that s$^{*}$ is quite high in energy, the second is that  V$_{s^{*}p\sigma}$ are the only hybridizations between s$^{*}$  and the sp$^3$ orbitals. Regarding the topological properties, these are not influenced at all by s$^{*}$. Indeed, at the $\Gamma$ point, the sp$^3$ energetic levels are unaffected by s$^{*}$ since the eigenvalues at $\Gamma$ are independent of V$_{s^{*}p\sigma}$. 
In the Appendix \ref{sec:appA}, we report how removing the s$^{*}$ orbital from the third model of Table I improves the agreement with the experimental results.
Experimentally, the energy of the $\Gamma_6$ state lies approximately 415~meV below $\Gamma_8$, as determined by ARPES and magneto-optical measurements~\cite{Polacinski2023}. The $\Gamma_7$ state is located about 800~meV below $\Gamma_8$. In our tight-binding model, $E(\Gamma_8) - E(\Gamma_7) = \Delta_0 = \tfrac{3}{2}\lambda = 681$~meV. 
The magnitude of the band inversion $E(\Gamma_8) - E(\Gamma_6)$ is 415~meV for our model, while for the third model in Table \ref{table:1} (with the inclusion of SOC), it is 425 meV.}

\begin{figure}[!t]
\includegraphics[width=1.0\columnwidth]{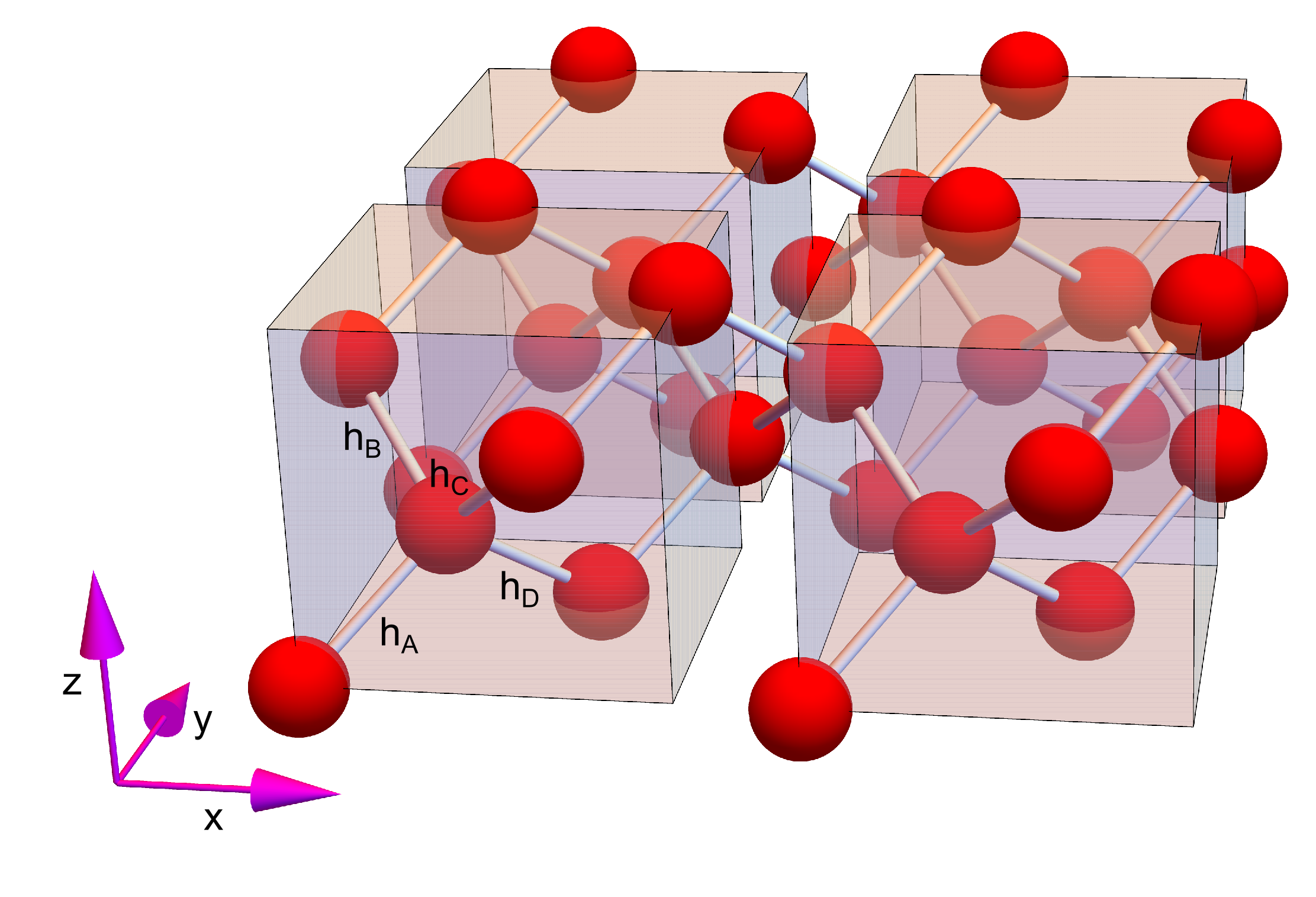}\caption{Schematic illustration of the crystal lattice of $\alpha$-Sn consisting of two layers of biatomic unit cells stacked along the $z$-direction. The Sn atoms are represented as balls and 
nearest-neighbor bonds $h_\alpha$ ($\alpha=A, B, C, D$)  as tubes connecting them. 
Each site supports one $s$-orbital and three $p$-orbitals. The shaded cubes indicate cubic unit cells.
\label{fig:alpha_tin_view}}
\end{figure}%figure2

\section{Tight-binding Hamiltonian}

\begin{figure}[!t]
\includegraphics[width=1.0\columnwidth]{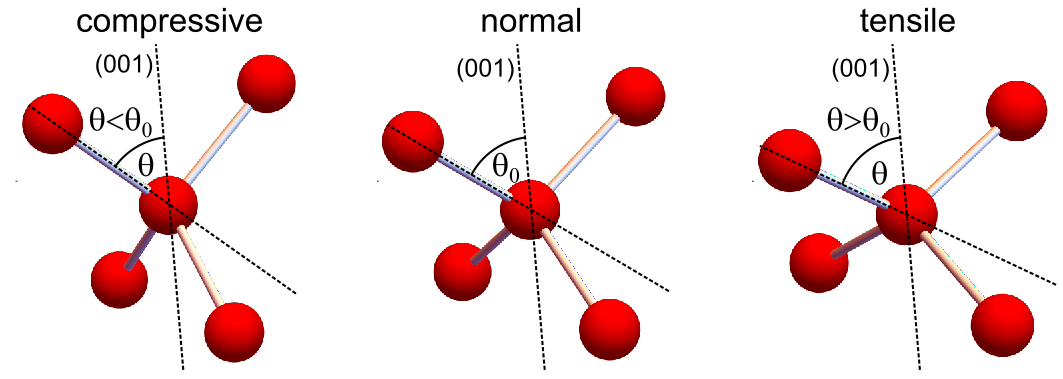}\caption{Schematic illustration of bonds of $\alpha$-Sn in case of compressive and tensile strain along $x$ and $y$ axes and in case without a strain. In the unstrained case, all bonds are equal, while in the strained case, there are two different sets of bonds. \label{fig:alpha_tin_strain}}
\end{figure}

Our starting point is a tight-binding Hamiltonian for $\alpha$-Sn. We consider a primitive unit cell consisting of two Sn atoms at positions $(0,0,0)$ and $(1/2,1/2,1/2)$ and lattice translation vectors being $\vec{n}_1=(1,1,0)$, $\vec{n}_2=(-1,1,0)$ and $\vec{n}_3=(0,1,1)$. The lattice structure is that of a diamond; see Fig. \ref{fig:alpha_tin_view}. 
Every Sn atom hosts one $s$ and three $p$ orbitals and we consider hopping only between nearest neighbors. We allow for all hopping processes that preserve spin and use the Slater-Koster parametrization\cite{Slater54} to write down the hopping matrix in the form of: % 
\begin{align}
h_{\left(l,m,n \right)} = \hspace{6.5 cm} \nonumber\\
\begin{pmatrix}
V_{ss\sigma} & lV_{s p\sigma} & mV_{s p\sigma}& nV_{s p\sigma}  \\
 -lV_{s p\sigma} &  l^2 \delta V+V_{p p\pi}&  lm \,\delta V&  ln\,\delta V \\
  -mV_{s p\sigma} &  lm \,\delta V  & m^2 \delta V+V_{p p\pi}  & mn \,\delta V\\
   -nV_{s p\sigma}&  ln \,\delta V&  mn \,\delta V&  n^2 \delta V+V_{p p\pi} 
   \end{pmatrix}.    
\end{align}
Where $\delta V=V_{p p\sigma}-V_{p p\pi}$ and where $(l,m,n)$ are direction cosines of a bond along which the electron hops. The parameters $V_{ss\sigma}$, $V_{pp\sigma}$ and $V_{pp\pi}$ are $\sigma$- or $\pi$-bonding amplitudes between two $s$ or $p$ orbitals, respectively. For NN hopping, we have four possible bond directions that depend on whether the system is strained or not. Here we parametrize the effect of in-plane tensile or compressive strain by introducing the bond angle $\theta$ depicted in Fig. \ref{fig:alpha_tin_strain}. Consequently, the  bond direction can be expressed by four normalized vectors $\vec A$, $\vec B$, $\vec C$ and $\vec D$ given by:  
\begin{eqnarray}
    \vec{A}_{\theta}=\hat{R}_{\theta}^{(\bar{1},1,0)}.\vec{e}_z,&\quad&\vec{B}_{\theta}=\hat{R}_{\pi-\theta}^{(1,1,0)}.\vec{e}_z,\nonumber\\
    \vec{C}_{\theta}=\hat{R}_{-\pi+\theta}^{(1,1,0)}.\vec{e}_z,&\quad&\vec{D}_{\theta}=\hat{R}_{-\theta}^{(\bar{1},1,0)}.\vec{e}_z,
\end{eqnarray}
where $\vec{e}_z=(0,0,1)^T$ and $\hat{R}_{\theta}^{\vec n}$ is a $\theta$-rotation operator with respect to the axis $\vec n$. In the case of an unstrained, fully symmetric diamond lattice, we have $\theta=\theta_0=\arctan\sqrt 2$, which yields  $\vec{A}=\sqrt 3^{-1}(1,1,1)$, $\vec{B}=\sqrt 3^{-1}(1,-1,-1)$, $\vec{C}=\sqrt 3^{-1}(-1,1,-1)$, $\vec{D}=\sqrt 3^{-1}(-1,-1,1)$. This angle decreases in the case of compressive strain and increases in the case of tensile strain. 
Having established the orbital hopping matrices, we can write the full tight-binding Hamiltonian as: 
\begin{eqnarray}
    \mathcal{H}_{\theta}(\textbf{k}) &\!=\!&\mathds{1}_2\!\otimes\!h_{\vec{A}_{\theta}}\!\!\otimes\!\begin{pmatrix}
0 & e^{ik_{3}} \\
 0&0 
\end{pmatrix} + \mathds{1}_2\!\otimes\!h_{\vec{B}_{\theta}}\!\!\otimes\!\begin{pmatrix}
0 & e^{ik_{1}} \nonumber\\
 0&0 
\end{pmatrix} \\
&\!+\!&\mathds{1}_2\!\otimes\!h_{\vec{C}_{\theta}}\!\!\otimes\!\begin{pmatrix}
0 & 1 \\
 0&0 
\end{pmatrix} + \mathds{1}_2\!\otimes\!h_{\vec{D}_{\theta}}\!\!\otimes\!\begin{pmatrix}
0 & e^{i\left( k_{3} -k_{2}\right)}\\
 0&0 
\end{pmatrix} \nonumber\\
&\!+\!& H.c. +  \mathds{1}_2 \!\otimes\! E_{\rm orb} \!\otimes\! \mathds{1}_2 +   \lambda\!\!\!\sum_{\alpha=x,y,z} \!\!\!   \sigma_{\alpha} \!\otimes\! \mathit{L}_{\alpha}\! \otimes\! \mathds{1}_2,
\label{eq:ham}
\end{eqnarray}
where $e^{i k_1}$, $e^{ik_2}$ and $e^{ik_3}$ are the eigenvalues of the lattice translation operators along lattice vectors $\vec{n}_1$,  $\vec{n}_2$ and  $\vec{n}_3$. To work with Cartesian quasimomenta, we can use a substitution $k_1=k_x+k_y$, $k_2=k_y-k_x$, $k_3=k_y+k_z$. 

In the on-site parts of the Hamiltonian, we find $E_{\rm orb}$ being a diagonal $4\times 4$ matrix, describing occupation energies of one $s$ and three $p$ orbitals, with diagonal entries given by $\{E_s,E_p,E_p,E_p\}$. Finally, there is also an atomic spin-orbit coupling with amplitude $\lambda$, where $\sigma_{\alpha}$ describes electron spin and $L_{\alpha}$ its orbital degree of freedom. Here, angular-momentum operators have a block-diagonal structure $L_{\alpha}=L^{(s)}\oplus L^{(p)}_{\alpha}$ with $L^{(s)}=0$ being the trivial angular momentum of the $s$ orbital and $L^{(p)}_{\alpha}$ being $L=1$ angular momentum of the $p$ orbitals acting on the basis of cubic harmonics $p_x$, $p_y$ and $p_z$.

The parameters $E_s$, $E_p$, $\lambda$, $V_{ss\sigma}$, $V_{pp\pi}$, $V_{ss\sigma}$ of the tight-binding Hamiltonian are given in 
Table~\ref{table:1} and have been derived from first-principles 
density-functional theory (DFT) calculations. 
To obtain these parameters with first-neighbour hopping parameters, we impose on the tight-binding model to fit the DFT band structure at high-symmetry points, extracting the on-site energies, the hopping amplitudes and the spin-orbit couplings as fitting parameters.

\begin{figure}[!t]
    \centering
    \includegraphics[width=0.995\linewidth]{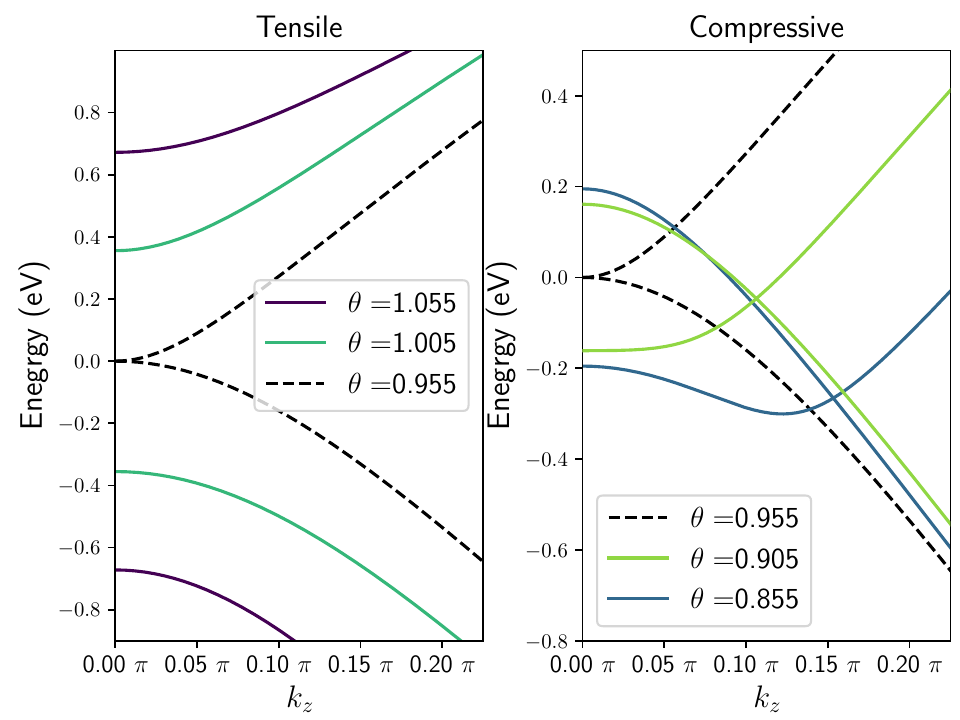}
    \caption{The response of the bulk band structure around the $\Gamma$-point to the tensile (left) and compressive (right) in-plane strain. This strain is modeled as a small change in the polar angle $\theta$ of the bond between the sublattices. The unstrained $\alpha$-Sn bond angle $\theta=\theta_0=0.955$ (black dashed line in both panels) serves as a reference. The increase/reduction of $\theta$ indicates in-plane stretching/compression of the sample. The comparison indicates that only a compressive strain can induce band inversion at the $\Gamma$-point.}
    \label{fig:pressure_vs_band_inversion}
\end{figure}

\section{Topological phase diagram and edge-states}

Tensile or compressive strain in the system is controlled by the angle $\theta$. In Fig. \ref{fig:pressure_vs_band_inversion} we show the band structure of the Hamiltonian (\ref{eq:ham}) as a function of $k_3$ for $k_1=k_2=0$. As expected from earlier studies \cite{Polacinski2023}, the tensile strain opens a gap at half-filling, whereas compressive strain produces a Dirac point. Therefore, we see that by controlling the strain, we will most likely be able to control the topological phase of the system. 

From the point of view of the tenfold Altland-Zirnbauer classification, our system is clearly in the AII class with spin-full time-reversal symmetry. In addition, it also has inversion symmetry ${\cal I}$ given by an operator    
\begin{equation}
    {\cal I}= \mathds{1}_2\!\otimes\![\mathds{1}_1\oplus(-\mathds{1}_3)]\!\otimes\!\tau_x.
\end{equation}
where $\tau_x$ is the $x$ Pauli matrix acting in the sublattice space.
From the work of Fu and Kane \cite{FuKane2007}, we know that in such a case, a $\mathbb{Z}_2$ invariant of the AII class can be calculated by looking at inversion eigenvalues of the occupied states at the high-symmetry points of the Brillouin zone (BZ), namely
\begin{equation}
  (-1)^\nu=\prod_{\vec K}\prod_{n\in N_f} \langle\psi_{\vec K}|{\cal I}|\psi_{\vec K}\rangle,
\end{equation}
where ${\vec K}$ are high-symmetry momenta and $N_f=\{1,3,\dots,7\}$ is a set of labels of the occupied states (one from each Kramers doublet, no matter which one) and $\nu=0,1$ for the trivial/non-trivial case, respectively. Using this prescription, we can calculate the invariant both in two and three spatial dimensions.

In the 3D case, we find that the system is always in a topological insulator phase with $\nu = 1$ as long as the tensile strain is applied. In the unstrained case, there is a parabolic band-touching at the $\Gamma$ point which evolves into two linear band-crossings present at $\vec{k}=(0,0,\pm \xi_z)$ for any compressive strain (see Fig. \ref{fig:pressure_vs_band_inversion}). Interestingly, while the system remains gapless in this limit, the $\nu$ invariant does not change with respect to the topological phase. Finally, we have also found that in the tensile strain case, the system realizes a topological crystalline insulator phase with a $(110)$-mirror Chern number $C=1$. This is a similar behavior to the one exhibited by SnTe-class compounds \cite{Story2012,Tanaka2012}, which suggests that the Dirac cones found at the surface of $\alpha$-Sn \cite{PhysRevB.98.195445} can survive breaking of the time-reversal symmetry as long as the mirror symmetry is preserved. In such a case, they would move away from the high-symmetry positions.

In the two-dimensional case, which is the main focus of this work, 
we are interested in a quantum-well configuration of $\alpha$-Sn, so we will stack a given number of layers of $\alpha$-Sn in the direction of $\vec{n}_3$.
\begin{figure}[!t]
    \centering
    \includegraphics[width=0.995\linewidth]{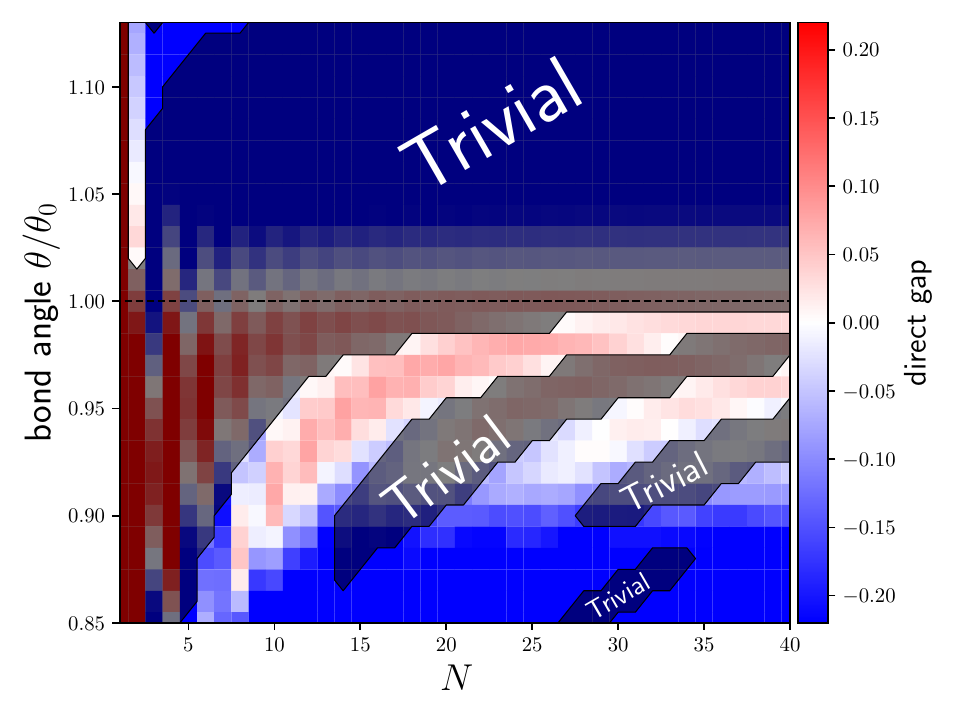}
    \caption{Phase diagram of a layered $\alpha$-Sn system. On the horizontal axis is $N$, the number of layers along the $\vec{n}_3$ direction. On the vertical axis is the bond polar angle between the sublattices divided by the unstrained angle $\theta_0$. Values larger (smaller) than one indicate tensile (compressive) in-plane strain. The color scheme denotes the value of the direct gap between the conduction and valence bands at the high-symmetry points. The blue and white regions are regions without a direct gap. The additional dark shading represents the regions where the system is topologically trivial.}
    \label{layers:topological_phase_diagram}
\end{figure}
The topological phase diagram as a function of the number of layers and the angle $\theta$ is given in Fig. \ref{layers:topological_phase_diagram}. 
As we can see, first of all, the system does not always stay insulating and this depends mainly on $\theta$. We clearly see the window of $\theta$ starting roughly at $0.9\theta_0$ and ending roughly at $1.01\theta_0$, where the system has a direct gap for any number of layers. Below $15$ layers, this window extends more to the lower values of $\theta$, apart from a single-layer case, which remains gapped for any angle $\theta$ considered here. The non-trivial phase appears for $\theta<\theta_0$, as long as there is more than one layer of the system, and it takes a slightly weird shape of finger-like fringes that get narrower as the number of layers grows, consistent with the 3D properties of the system.

\begin{figure}
    \centering
    \includegraphics[width=0.995\linewidth]{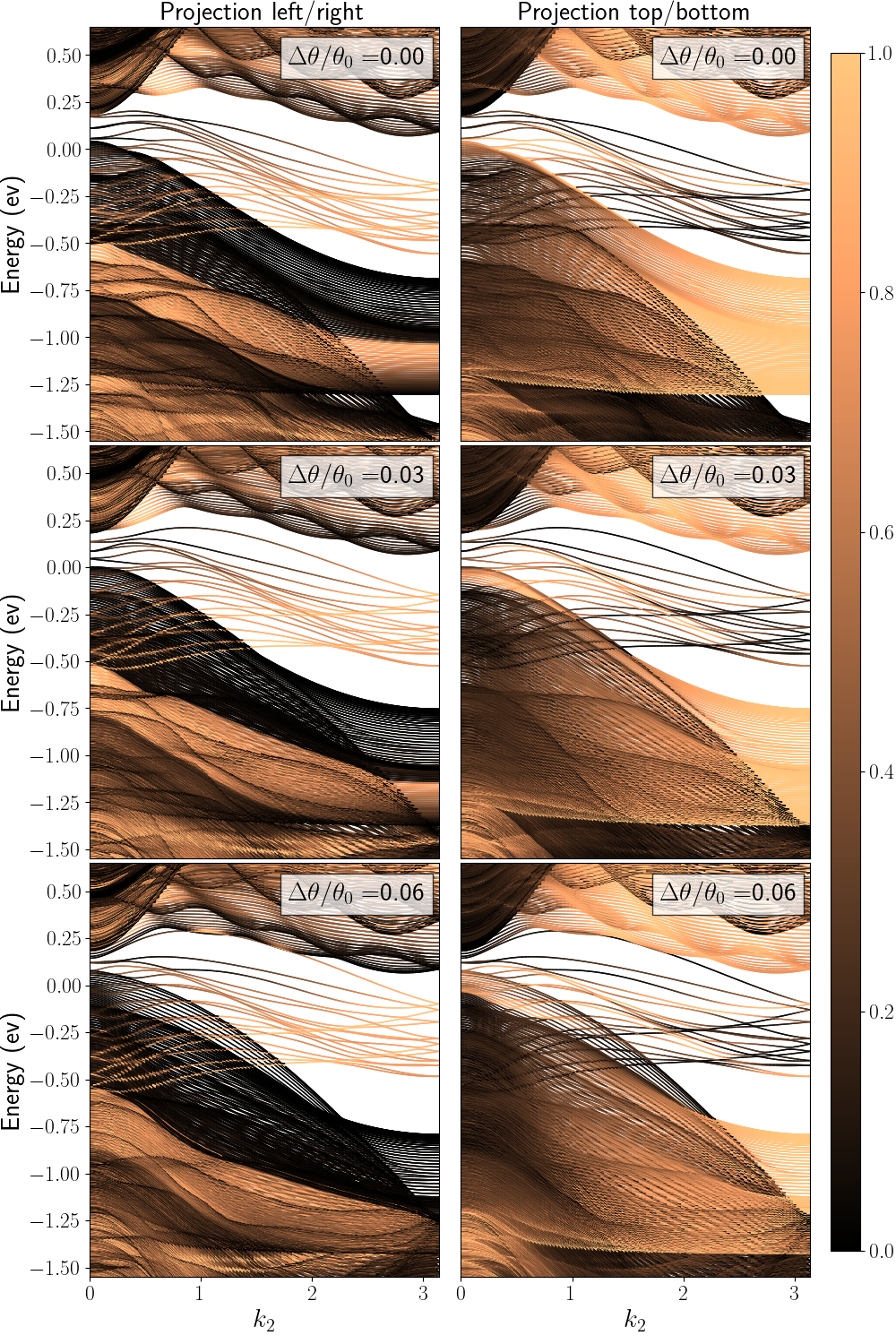}
    \caption{Band structure of a slab with $N_z=10$ and $N_x=80$, with the colors representing the projection of each band on the left/right (left) and top/bottom (right) surfaces of the slab. In each row, the results for a certain compressive strain ($\Delta\theta=\theta_0-\theta$), denoted in the top right corner, are shown. The same color scale, explained on the right side of the figure, is used in each panel. }
    \label{stack:bands_projection}
\end{figure}

Having established the topological phase diagram, we can now examine the edge states. In Fig. \ref{stack:bands_projection}, we show the band structures around the Fermi level in the case of $10$ layers of the system with open edge parallel to the $\vec{n}_2$ direction. We present the reference case with no strain, as well as small compressive and larger compressive strains. However, what is interesting here is that according to the phase diagram, we should only see the in-gap edge-states in the last case, but we observe them in all cases. Looking at the spatial projections of these states, we clearly see that they are all strongly localized at the edges, especially for $k_2=\pi$, where the bulk gap is the largest (some of the states also prefer to stay more on top/bottom layers rather than in the central ones). We also note that these states are typically not connecting the valence and conduction bands - only in the larger-compressive-strain case, which is supposed to support the $\mathbb{Z}_2$ invariant, we see a pair of helical states which connect through the bulk gap.

Summarizing, the situation with the edge-state here is similar to the one encountered in the HgTe/CdTe quantum wells \cite{Brze2023}; however, we are dealing with an inversion-symmetric system that is also homogeneous along the stacking direction.

\section{Hidden winding numbers}

\begin{figure}[!t]
\includegraphics[width=1.0\columnwidth]{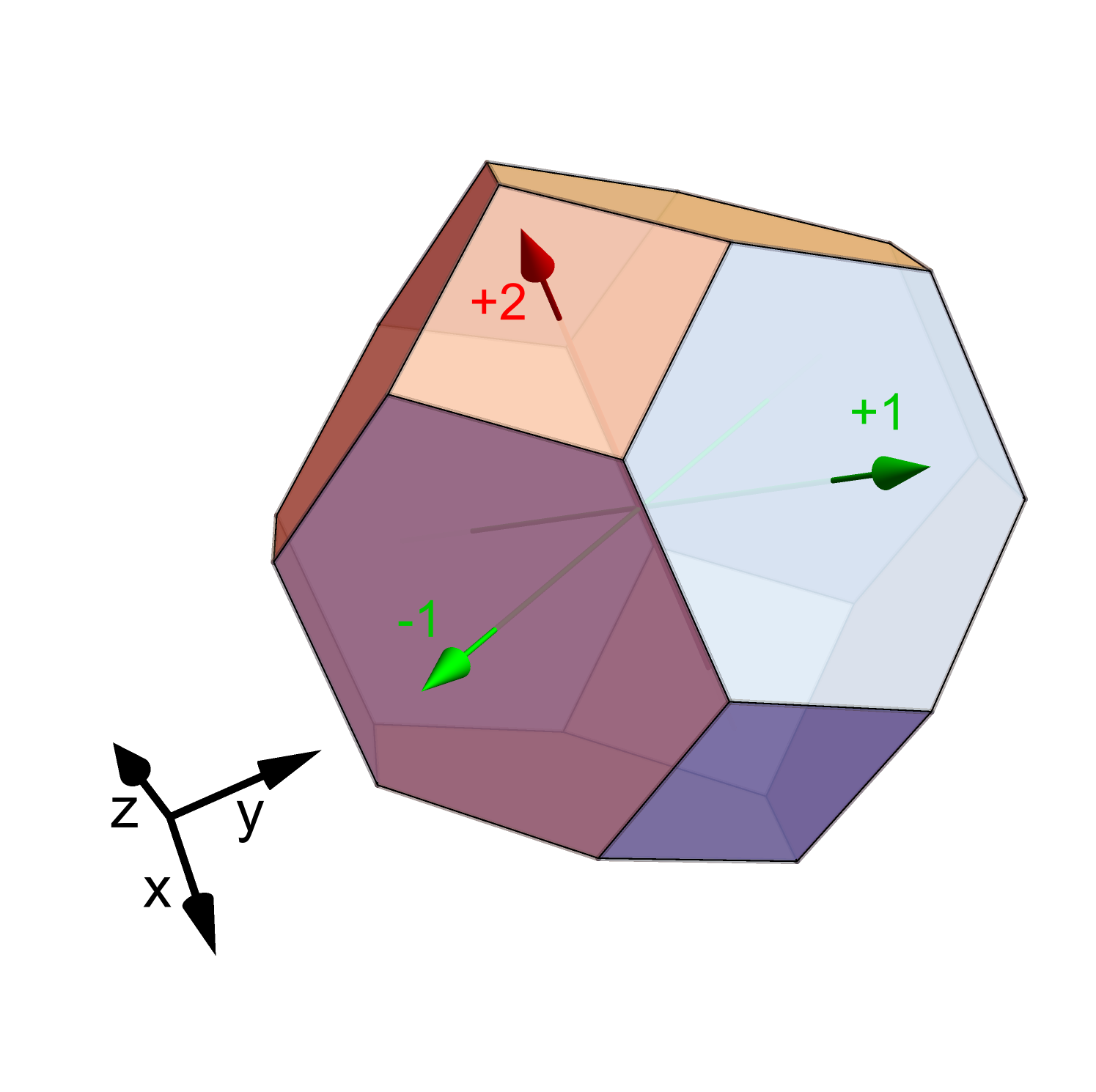}\caption{Diamond-lattice Brillouin zone with marked 1D winding numbers $\nu_{1\rm D}$ calculated along directions indicated by red and green arrows.  \label{fig:BZ_nu}}
\end{figure}

Our task now is to understand the origin of the additional edge-states shown in Fig. \ref{stack:bands_projection}. This can be achieved by continuously evolving the Hamiltonian to a point where it has more symmetries. Based on an educated guess, given earlier works showing that the diamond lattice can support 1D winding numbers \cite{Takahashi2013}, we choose the direction of this evolution to be towards a chiral-symmetric system. This can be obtained by setting:
\begin{equation}
    E_p = E_s,\quad \lambda = 0,
\end{equation}
and subtracting the constant energy shift of $E_s$, i.e., 
\begin{equation}
   {\cal H}_{\rm chir}(\textbf{k}) = {\cal H}_{\theta_0}(\textbf{k},E_p = E_s, \lambda = 0) - E_s \mathds{1}_{16}.
\end{equation}
Note that in this Section, we will focus on the unstrained case of $\theta = \theta_0$ for the sake of simplicity, but the reasoning stays valid for any other $\theta$. 
From the fact that the diamond lattice is bipartite and the Hamiltonian contains only NN hopping, it is easy to infer that ${\cal H}_{\rm chir}(\textbf{k})$ has a sublattice or chiral symmetry given by 
\begin{equation}
  {\cal S} = \mathds{1}_2 \!\otimes\! \mathds{1}_4 \!\otimes\! \begin{pmatrix}
1 & 0 \\
0 & -1 
\end{pmatrix},
\end{equation}
satisfying $\{{\cal H}_{\rm chir}(\textbf{k}), {\cal S}\}\equiv0$. From this it follows that we can block-off-diagonalize ${\cal H}_{\rm chir}(\textbf{k})$ in the eigenbasis of ${\cal S}$. Additionally, since we have turned off the spin-orbit coupling, the spin channels are now decoupled, so we can consider only the spin-up Hamiltonian. Thus, we have:
\begin{equation}
{\cal H}^{\uparrow}_{\rm chir}(\textbf{k})=
\begin{pmatrix}
0 & u_{\textbf{k}} \\
u_{\textbf{k}}^{\dagger} & 0 
\end{pmatrix},
\end{equation}
where 
\begin{equation}
u_{\textbf{k}}=
\begin{pmatrix}
q^{(1)}_{\textbf{k}}V_{ss\sigma} & 
q^{(2)}_{\textbf{k}}\frac{V_{sp\sigma}}{\sqrt{3}} & 
-q^{(3)}_{\textbf{k}}\frac{V_{sp\sigma}}{\sqrt{3}} & 
-q^{(4)}_{\textbf{k}}\frac{V_{sp\sigma}}{\sqrt{3}}  \\
-q^{(2)}_{\textbf{k}}\frac{V_{sp\sigma}}{\sqrt{3}} & 
q^{(1)}_{\textbf{k}}\frac{V_{+}}{3} & 
q^{(4)}_{\textbf{k}}\frac{V_{-}}{3} & 
q^{(3)}_{\textbf{k}}\frac{V_{-}}{3} \\
q^{(3)}_{\textbf{k}}\frac{V_{sp\sigma}}{\sqrt{3}} & 
q^{(4)}_{\textbf{k}}\frac{V_{-}}{3} & 
q^{(1)}_{\textbf{k}}\frac{V_{+}}{3} & 
-q^{(2)}_{\textbf{k}}\frac{V_{-}}{3} \\
q^{(4)}_{\textbf{k}}\frac{V_{sp\sigma}}{\sqrt{3}} & 
q^{(3)}_{\textbf{k}}\frac{V_{-}}{3} & 
-q^{(2)}_{\textbf{k}}\frac{V_{-}}{3} & 
q^{(1)}_{\textbf{k}}\frac{V_{+}}{3} 
\end{pmatrix},
\end{equation}
and
\begin{equation}
V_{+} = 2V_{pp\pi}+V_{pp\sigma},\quad V_{-} = V_{pp\pi}-V_{pp\sigma},
\end{equation}
as well as:
\begin{eqnarray}
    q^{(1)}_{\textbf{k}} &=& 1 + e^{ik_1} + e^{ik_3} + e^{i(k_3-k_2)},\nonumber\\
    q^{(2)}_{\textbf{k}} &=& -1 + e^{ik_1} + e^{ik_3} - e^{i(k_3-k_2)},\nonumber\\
    q^{(3)}_{\textbf{k}} &=& -1 + e^{ik_1} - e^{ik_3} + e^{i(k_3-k_2)},\nonumber\\
    q^{(4)}_{\textbf{k}} &=& 1 + e^{ik_1} - e^{ik_3} - e^{i(k_3-k_2)}.\nonumber\\
\end{eqnarray}
\begin{figure}[!t]
    \centering
    \includegraphics[width=0.995\linewidth]{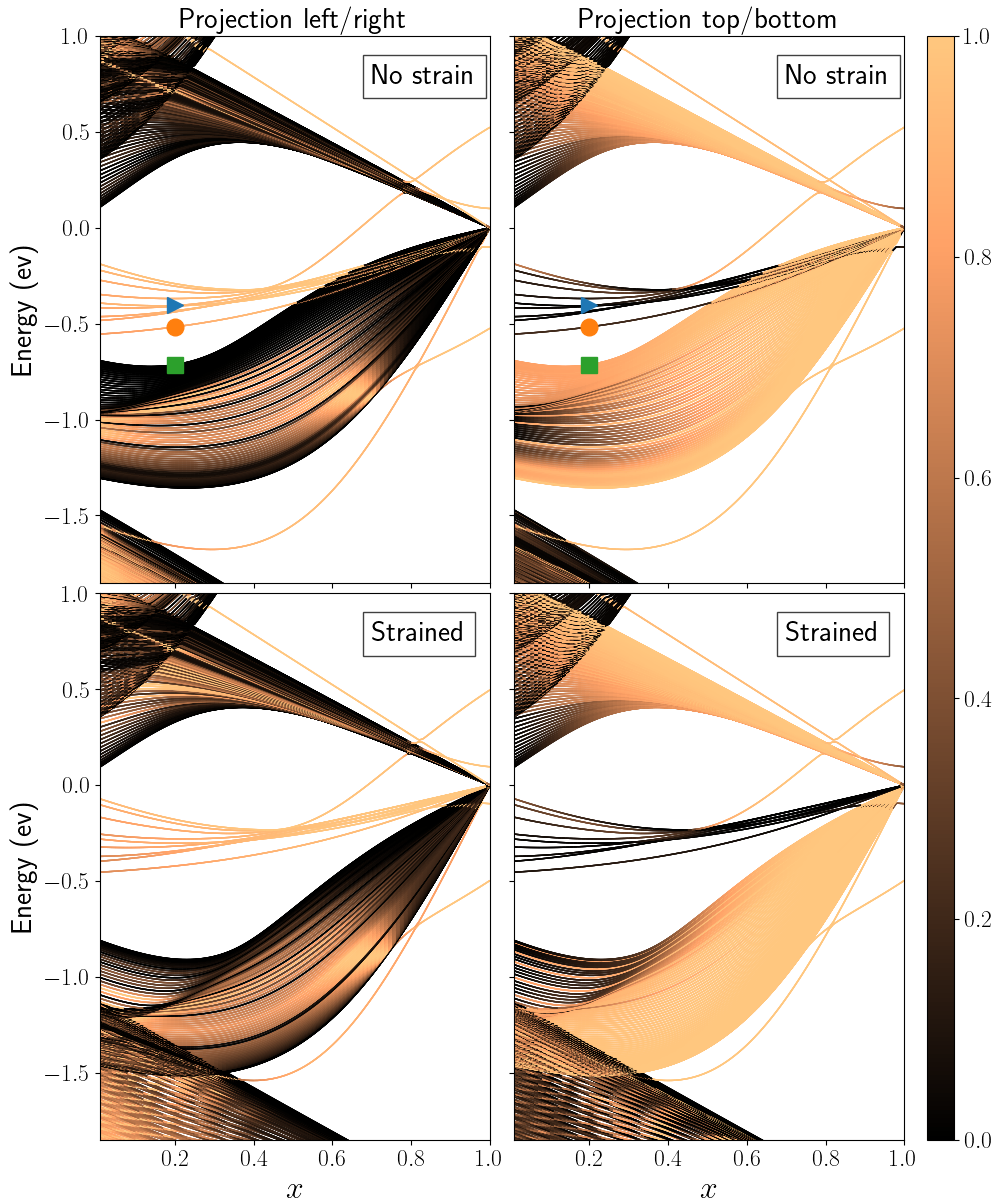}
    \caption{Band structure at $k_2=\pi$ of unstrained (top) and strained (bottom) slab of $\alpha$-tin model with $N_3=10$ and $N_1=80$ as the parameters evolve from the experimental values ($x=0$) to the chiral effective model ($x=1$). The color scale, explained on the right side, describes the projection of the spectral weight of each band at the left/right (left column) and top/bottom (right column) termination surfaces. The blue triangles, orange circles, and green squares denote specific in-gap states that will be examined later.}
    \label{stack:interpolation}
\end{figure}
The Hamiltonian ${\cal H}^{\uparrow}_{\rm chir}$ is clearly of the BDI class, so it can support a chiral $\mathbb{Z}$ invariant in three dimensions or weaker chiral $\mathbb{Z}$ invariants (winding numbers) living in the 1D cuts of 3D BZ. It turns out that the latter case is the one that is indeed realized and it is quite intriguing. We find that the values of the winding numbers depend on the direction in the BZ. In Fig. \ref{fig:BZ_nu} we show the diamond-lattice BZ with different winding numbers $\nu_{1\rm D}$ indicated depending on the chosen direction (note that we use Cartesian $k_{x,y,z}$ here). We observe that if we decide to go along the four-fold rotation axes, we always get $\nu_{1\rm D}=2$, but if we go along the six-fold rotation axes, it takes values $\pm 1$ depending on which axis we choose. This is clearly related to the fact that different ending surfaces can either consist of atoms belonging only to one sublattice (in the case of six-fold rotation-invariant surfaces) or to both (in the case of four-fold rotation-invariant surfaces).
\begin{figure}[!t]
    \centering
    \includegraphics[width=0.995\linewidth]{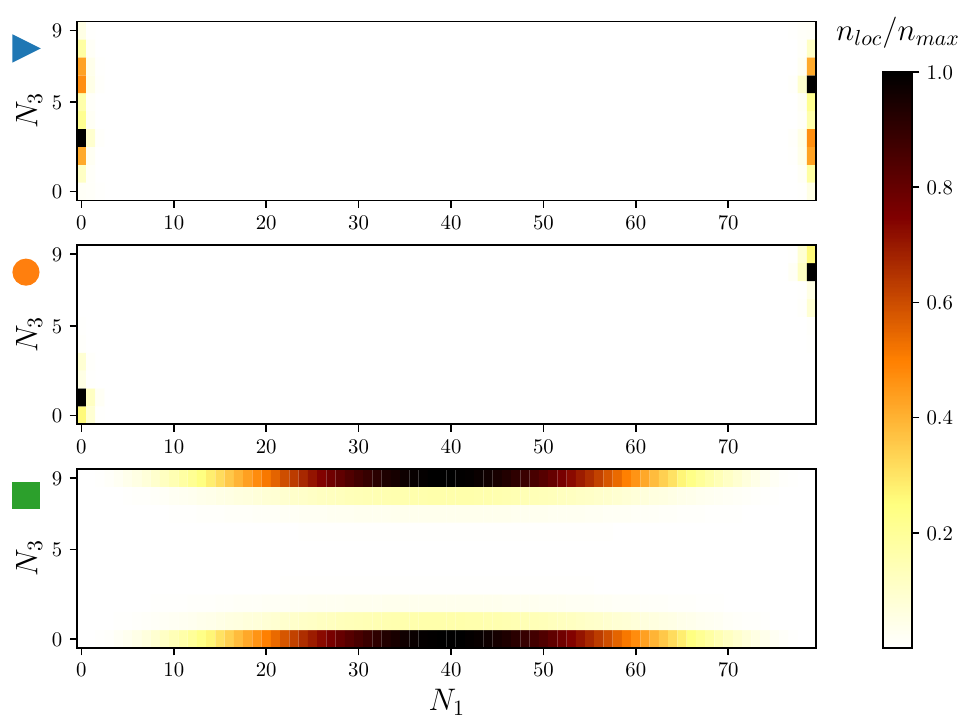}
    \caption{Local density of electrons (normalized to the maximal value) across the unstrained slab of $N_3=10$ and $N_1=80$ and parameter $x=0.2$ for various in-gap states at $k_2=\pi$. The in-gap states are indicated by blue triangles, orange circles, and green squares in Fig. \ref{stack:interpolation}.}
    \label{stack:local_densities}
\end{figure}
Finally, to prove the relevance of the chiral model to the original one, we show in Fig. \ref{stack:interpolation} the evolution of the band structure at $k_2=\pi$ of a system with an open edge upon linear interpolation between the full and the chiral symmetric Hamiltonian. The interpolation parameter is $x$ and for $x=0$ we have a full model and for $x=1$ the chiral effective one, so with $E_p = E_s$ and $\lambda = 0$. We see that our edge-states smoothly evolve into a flat band at $E=0$ whose existence follows from a non-zero winding number. Note that some of the valence and conduction states also join the flat band and these are the ones that are localized at the top/bottom surfaces. This follows from the fact that both left/right and top/bottom surfaces have related winding numbers that lead to zero-energy states localized at these surfaces. This also allows us to predict how many additional edge-states we get in the original system, depending on the number of layers.

The spatial profile of different kinds of edge-states shown in Fig. \ref{stack:interpolation} is given in Fig. \ref{stack:local_densities}. 
Quite surprisingly, we still see that apart from states localized on the left/right or top/bottom surfaces, we also get some resembling hinge-states. In the case of the slab considered here, however, the lack of any spatial symmetry excludes the existence of rigorous higher-order topological states. We suspect that their presence is related to the mirror Chern number, like that of SnTe \cite{Schindler18}, which can lead to topological hinge states when the hinge is compatible with mirror symmetry,
as discussed in the next Section.
We also note that these states do not join the flat band at $x=1$, so they are not related to the winding numbers. 
\section{Role of the mirror symmetry}
\begin{figure}
   \centering
   \includegraphics[width=1.0\linewidth]{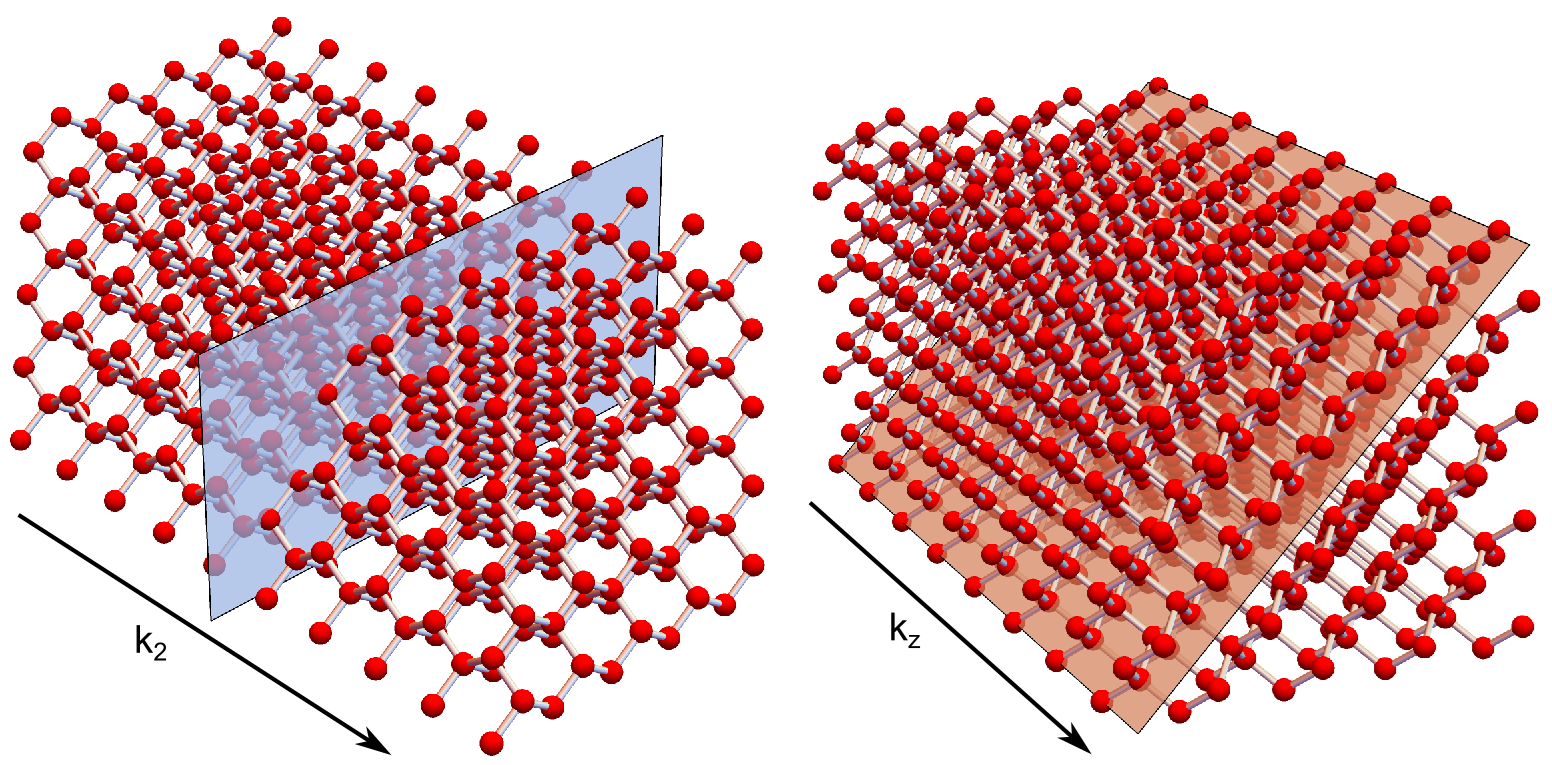}
    \caption{Schematic view of the systems being finite in the directions of $\vec{n}_1$ and $\vec{n}_3$ (left) or in the directions $x$ and $y$ (right). The relevant mirror symmetry planes are indicated. The systems remain translationally invariant in the directions $\vec{n}_2$ (left) and $z$ (right), hence $k_2$ and $k_z$.}
   \label{fig:wire_views}
\end{figure}

Mirror symmetries play an interesting role in the case of tensile strain. For the Hamiltonian (\ref{eq:ham}), we find two mirror operators which are compatible with the system being finite along the $\vec{n}_3$ direction, namely the ones with mirror planes being $(110)$ and $(1\bar{1}0)$. These are defined as:
\begin{eqnarray}
          \mathcal{M}_{1}(k_1)\!\!&=&\!\frac{\sigma_x\!-\!\sigma_y}{\sqrt 2}\!\otimes\!\left[(L_x-L_y)^2\!-\!\vec{L}^2\!+\!\mathds{1}_4\right]\!\otimes\!\begin{pmatrix}
1 & 0 \\
0 & e^{ik_{1}} 
\end{pmatrix},\nonumber\\
      \mathcal{M}_{2}\!\!&=&\!\frac{\sigma_x\!+\!\sigma_y}{\sqrt 2}\!\otimes\!\left[(L_x+L_y)^2\!-\!\vec{L}^2\!+\!\mathds{1}_4\right]\!\otimes\!\mathds{1}_2.
\end{eqnarray} 
Their relation to the Hamiltonian is given by:
\begin{eqnarray}
    \mathcal{M}_{1}(k_1)\mathcal{H}_{\theta}(k_1,k_2,k_3)\mathcal{M}_{1}^{\dagger}(k_1) &\!=\!&\mathcal{H}_{\theta}(-k_1,k_2,k_3-k_1), \nonumber \\
    \mathcal{M}_{2}\mathcal{H}_{\theta}(k_1,k_2,k_3)\mathcal{M}_{2}^{\dagger} &\!=\!&\mathcal{H}_{\theta}(k_1,-k_2,k_3-k_2).\nonumber
\end{eqnarray}
In Fig. \ref{fig:wire_views} we show real-space views of the $\alpha$-Sn crystal lattice with different termination surfaces and one of these mirror planes explicitly marked. For a system composed of primitive unit cells (finite in $\vec{n}_{1,3}$ directions and periodic in $\vec{n}_2$), we see that each termination surface is mapped onto itself by the mirror symmetry and that quasimomentum $k_2$ is flipped under its action. On the other hand, for a system composed of cubic unit cells (shown in Fig. \ref{fig:alpha_tin_view}), we see that the adjacent termination surfaces are mapped onto each other by the mirror and the quasimomentum $k_z$ is unchanged.

\begin{figure}
   \centering
   \includegraphics[width=1.0\linewidth]{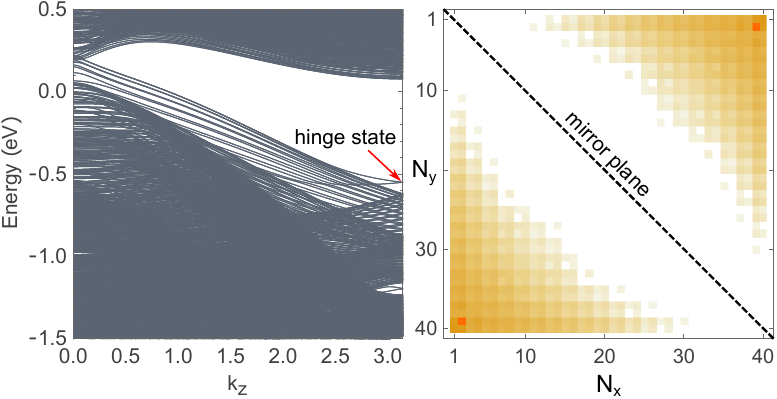}
    \caption{Band structure of the system being finite in the directions $x$ and $y$ and periodic along $z$ (left) of the size  $N_x=N_y=40$ primitive unit cells. On the right, the local charge density in the hinge-state manifold is indicated with a red arrow in the spectrum. The system is under tensile strain $\theta = 1.005$.}
   \label{fig:spec_hinges}
\end{figure}

In the latter configuration, it is not uncommon to get hinge states, as long as the system supports mirror Chern numbers \cite{Schindler18}, because hinges can be seen as edges of the mirror plane. This is exactly the case in our system under tensile strain. In Fig. \ref{fig:spec_hinges} we show a spectrum of the system with cubic termination surfaces. A pair of helical states is clearly visible in the gap close to $k_z$. It coexists with a bunch of other in-gap states, which appear in a similar manner as in the earlier discussed cases, but seem to connect through the gap. The charge distribution in these states is clearly localized at the hinges of the system, as shown in Fig. \ref{fig:spec_hinges}. Note that the energy of these hinge states at $k_z=\pi$ is very close to the energy of the state marked with an orange dot shown in Figs. \ref{stack:interpolation} and \ref{stack:local_densities}. 

\begin{figure}
   \centering
   \includegraphics[width=1.0\linewidth]{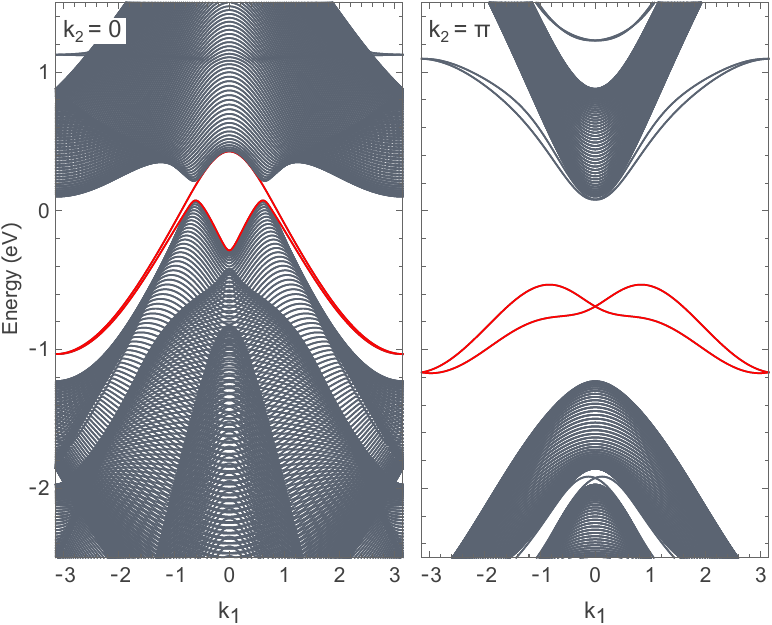}
    \caption{Band structures of the system being periodic in the directions of $\vec{n}_1$ and $\vec{n}_2$ and finite in $\vec{n}_3$ and consisting of $N_3=100$ layers. Spectra plotted versus $k_1$ for 
    $k_2=0$ (left) and $k_2=\pi$ (right). The topological in-gap states are shown in red. The system is under tensile strain $\theta = 1.005$.}
   \label{fig:edge_spec}
\end{figure}

Now we focus on the system composed of primitive unit cells, described by the Hamiltonian (\ref{eq:ham}). At high-symmetry points $k_2=0,\pi$ we have:
\begin{eqnarray}
        \mathcal{M}_{2}\mathcal{H}_{\theta}(k_1,0,k_3)\mathcal{M}_{2}^{\dagger} \!&=&\!\mathcal{H}_{\theta}(k_1,0,k_3),\nonumber\\
        \mathcal{M}_{2}\mathcal{H}_{\theta}(k_1,\pi,k_3)\mathcal{M}_{2}^{\dagger} \!&=&\!\mathcal{H}_{\theta}(k_1,\pi,k_3+\pi).
\end{eqnarray}
This means that for $k_2=0$, mirror symmetry $\mathcal{M}_{2}$ commutes with the Hamiltonian and indeed it yields a mirror Chern number $C=1$ in the case of tensile strain. On the other hand, for $k_2=\pi$, we get a rather peculiar commutation relation with an additional shift by $\pi$ in quasimomentum $k_3$. 
Such property is often called a momentum-space nonsymmorphic symmetry. 
This has an interesting impact on the band structure. In Fig. \ref{fig:edge_spec} we show edge-state spectra at $k_2=0$ and $k_2=\pi$ points as functions of $k_1$ for a system open in the $\vec{n}_3$ direction under tensile strain. For $k_2=0$, we see a pair of helical edge-states connecting through the gap. This is expected, as they follow from the mirror Chern number - the $\mathcal{M}_{2}$ operator commutes with the Hamiltonian, splits it into two diagonal blocks and they both carry opposite Chern numbers $\pm 1$. On the other hand, for $k_2=\pi$, the same symmetry does not split the Hilbert space into two separate pieces, because the relation with the Hamiltonian involves the $\pi$-shift. Nevertheless, we still find a pair of helical in-gap states, although they do not connect through the gap, suggesting that they follow from 1D topology. Indeed, making use of the $k_3\to k_3+\pi$ shift symmetry at $k_2=\pi$ we can define for every $k_1$-point a Wilson loop operator:
\begin{equation}
    \mathcal{W}^{n,m}_{k_1,\pi}=\langle \psi^n_{(k_1,\pi,-\pi)}|\left[\overrightarrow{\prod_{k_3=-\pi}^0}\mathcal{P}^{occ}_{(k_1,\pi,k_3)}\right] \mathcal{M}_2|\psi^m_{(k_1,\pi,-\pi)}\rangle.
\end{equation}
Here $n,m$ indices run over occupied states and $\mathcal{P}^{occ}$ are projectors on occupied states for every $k$-point. They are multiplied in the order of increasing $k_3$.
\begin{figure}
   \centering
   \includegraphics[width=0.9\linewidth]{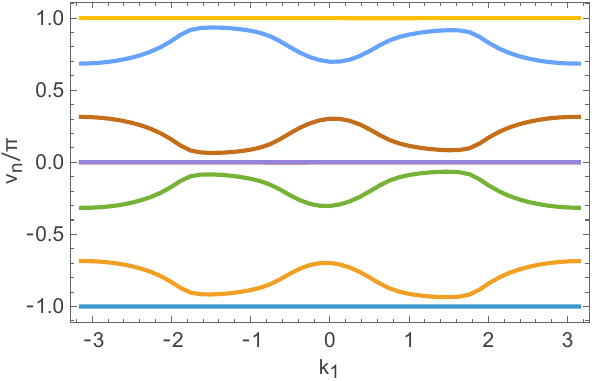}
    \caption{
    Phases of the eigenvalues of the Wilson-loop operators calculated along $k_3$ as functions of $k_1$ for $k_2=\pi$. The system is under tensile strain $\theta = 1.005$.}
   \label{fig:phases_pi}
\end{figure}
Note that the loop is done only over the non-positive $k_3$'s because the positive ones are related by the $\pi$-shift operation. The Wilson loop operators are unitary, so the eigenvalues $\mathcal{W}$ are all phase factors of the form $e^{i\nu_n}$. In Fig. \ref{fig:phases_pi} we show the evolution of these $8$ phases as functions of $k_1$. We see that two of them are degenerate at $0$, two are $\pm\pi$ and we also find two pairs of opposite phases. This is the same behavior that was discussed in Ref. \onlinecite{Alexandra14} in the context of the inversion topological invariant in 1D. Therefore, we argue that the number of $-1$ eigenvalues of $\mathcal{W}$ is a topological invariant here that yields the disconnected edge-states found in Fig. \ref{fig:edge_spec}.   

These two rigorous topological results establish another mechanism for appearance of additional edge-states at $k_2=0$ and $k_2=\pi$, similar to ones shown in Fig. \ref{stack:bands_projection}. Note that the mirror Chern number is non-vanishing only for tensile strain, so it will yield additional edge states at $k_2=0$ only in this limit. However, the Wilson loop invariant found at $k_2=\pi$ is quite independent from strain. This is because strain closes and opens the gap at the $\Gamma$ point whereas the Wilson loop-invariant lives at $k_2=\pi$ (or equivalently at $k_1=\pi$).
Indeed, we have found that Wilson-loop operators have two $-1$ eigenvalues also in the compressive and no-strain limits.

\section{Conclusions}

We have shown within a simple tight-binding model that multilayers of $\alpha$-Sn can host a QSH phase when compressive
strain is applied and this phase depends on the number of layers.
Similarly, as discovered in HgTe/CdTe quantum wells \cite{Brze2023}, both in the trivial and non-trivial phases, we obtain additional edge states within the energy gap when the edges of the 2D multilayer system are open.
Such states could lead to multimode edge-transport, distorting the sought-after quantum spin Hall effect.

By extrapolating the model to the chiral-symmetric limit we were able to assign the origin of 
these additional edge-states to the non-vanishing winding numbers present in the 1D cuts of the 3D 
BZ. The chiral-symmetric limit is reached by setting spin-orbit coupling to zero and removing 
crystal-field splitting between $s$ and $p$ orbitals.
When it is done the additional edge-states become zero-energy flat bands localized on top/bottom
or the left-right surfaces of the slab, depending on a particular value
of the winding number in a given direction.

The additional edge-states can also be  related to the the mirror symmetry. We have shown that those appearing at the $\Gamma$ point follow from the mirror Chern number in the tensile-strain limit whereas those at the zone boundary follow from the Wilson-loop invariant, no matter the strain. On the other hand, the mirror Chern number can also lead to topological hinge-states in case of cubic termination surfaces.  
%dual topology
The presence of the $\mathbb{Z}_2$ invariant and a non-zero mirror Chern number makes the present system one of the rare cases where dual topology can be observed.

The transition from a Dirac semimetal to the QSH phase 
when the system's dimensionality is reduced from three to two, 
could also be realized in other material systems.
Since thickness-dependent investigation of $\alpha$-Sn poses experimental challenges, similar physics might be more accessible in van der Waals materials, where exfoliation techniques can be readily applied. In this context, we identify PtSe$_2$ and PtTe$_2$ (space group No. 164) as promising candidates to host the QSH phase. These compounds have been reported to exhibit a 3D topological-insulator phase in the bulk and to evolve into normal insulators in the monolayer limit.\cite{Yan2017,PhysRevB.94.121117,PhysRevB.96.125102} Our theoretical results further suggest that intermediate-thickness films of these materials could host the QSH phase.

\vspace{1cm}
%\newpage

\section*{Acknowledgments}

We thank D. Zgid, T. Dietl, J. Polaczy\'nski, A. Kazakov and V. V. Volobuiev for their useful discussions. This research was supported by the Foundation for Polish Science project “MagTop” no. FENG.02.01-IP.05-0028/23 co-financed by the European Union from the funds of Priority 2 of the European Funds for a Smart Economy Program 2021–2027 (FENG). G.C. and C.A. acknowledge support from PNRR MUR project PE0000023-NQSTI.
W.B., J.S., N.M.N. and A.K. acknowledge support by Narodowe Centrum Nauki (NCN, National Science Centre, Poland) Project No. 2019/34/E/ST3/00404.
We acknowledge the access to the computing facilities of the Interdisciplinary Center of Modeling at the University of Warsaw, Grant G84-0, GB84-1 and GB84-7. We acknowledge the CINECA award under the ISCRA initiative  IsC85 "TOPMOST" and IsC93 "RATIO" grant, for the availability of high-performance computing resources and support. We acknowledge the access to the computing facilities of the Poznan Supercomputing and Networking Center, Grant No. 609.

The data and scripts to reproduce figures presented in
this manuscript are available at Ref.~\cite{zenodo}.

\appendix

\begin{figure}[!h]
\centering
\includegraphics[width=4cm,angle=0]{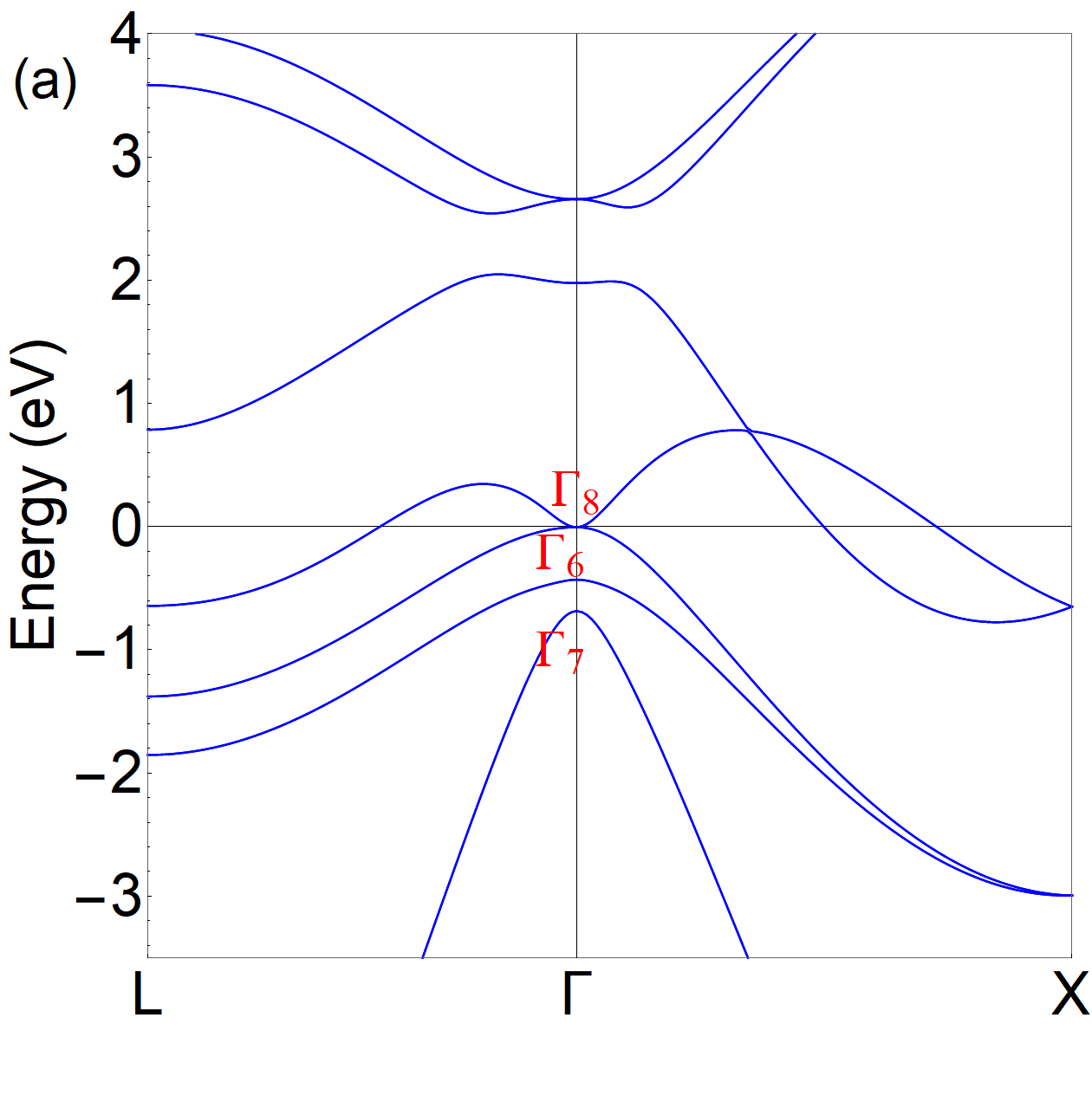}
\includegraphics[width=4cm,angle=0]{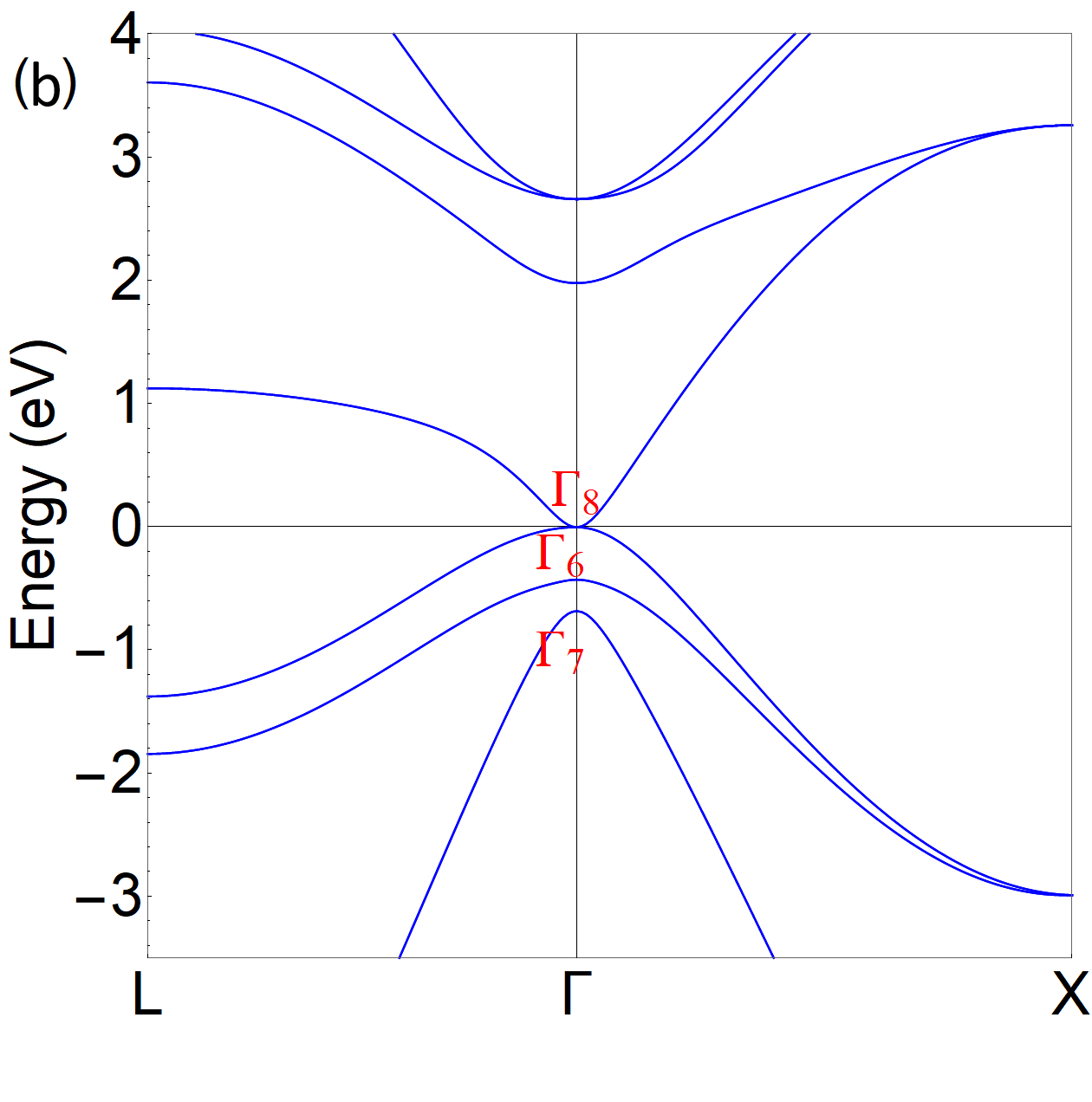}
\caption{Relativistic band structure of $\alpha$-Sn using the tight-binding model reported in Table \ref{table:1}. (a) Band structure in the energy range from -3.5 eV to +4 eV. (b) Band structure in an extended energy range. The X-point is at the coordinates ($\frac{1}{2}$,0,0), while the L-point is at the coordinates ($\frac{1}{2}$,$\frac{1}{2}$,$\frac{1}{2}$). $\Gamma_8$ and $\Gamma_7$ represent the p-states at the $\Gamma$ point while $\Gamma_6$ represents the s-state.
}
\label{Bands_tight_binding_appendix}
\end{figure} %figure10

\section{Removing s$^{*}$ form the tight-binding models \label{sec:appA}}

{In this Appendix, we examine the modification of the band structure upon removing the $s^{*}$ orbital from the third tight-binding model listed in Table~\ref{table:1}. As discussed in the main text, the s orbital does not affect the topological properties; however, it does influence the metallicity, as we will show here. When the $s^{*}$ orbital is included, we reproduce the band structure reported in the literature, shown in Fig.~\ref{Bands_tight_binding_appendix}(a). This dispersion exhibits metallic behavior along both the $\Gamma$--L and $\Gamma$--X directions, corresponding to a non-topological metal and thus conflicting with recent experimental results~\cite{Polacinski2023}.
Upon eliminating the $s^{*}$ orbital, the resulting band structure, displayed in Fig.~\ref{Bands_tight_binding_appendix}(b), instead describes a zero-gap topological semiconductor. We were unable to reproduce the results reported for the second tight-binding model in Table~\ref{table:1}, likely due to the use of a different tight-binding convention in that work.}

\section{{Suppression of trivial edge states by a surface potential}}

\begin{figure}[!h] 
\centering 
\includegraphics[width=0.99\linewidth]{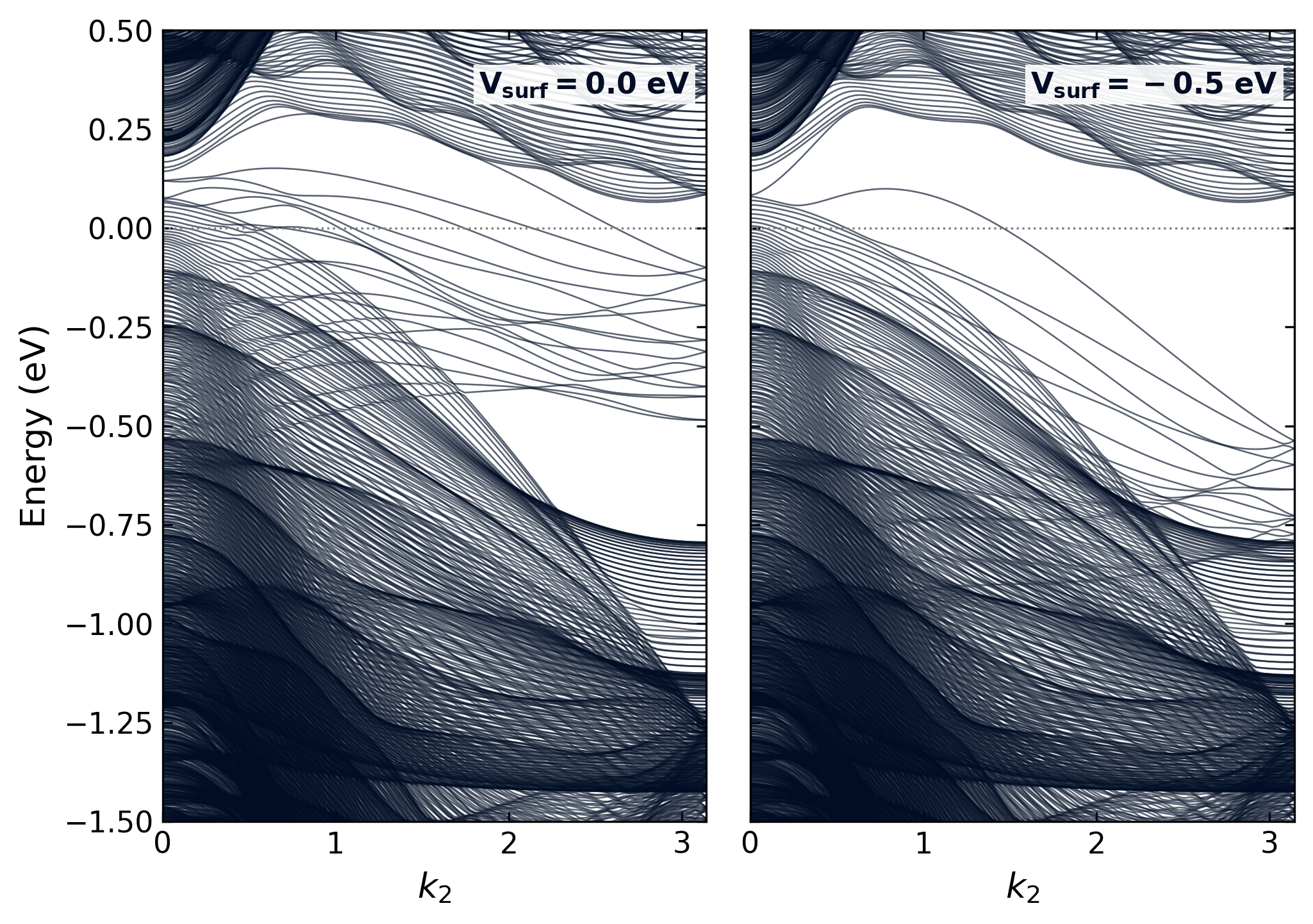} 
\caption{ Effect of a surface potential on the edge-state spectrum.} \label{fig:surface_potential_spectrum} 
\end{figure}

{To investigate possible strategies for suppressing the additional in-gap states, we introduce a surface potential $V_{\mathrm{surf}}$ acting on the outermost surface layers.  Figure~\ref{fig:surface_potential_spectrum} compares the band structure in the absence of a surface potential and for $V_{\mathrm{surf}}=-0.5$ eV. }

{Without a surface potential, numerous in-gap states coexist with the topological edge modes. Upon applying a sufficiently strong negative surface potential, these additional states are shifted into the bulk bands. This demonstrates that surface engineering provides a practical route for suppressing unprotected edge channels and improving the observability of quantized transport associated with the topological edge states.}

\section{{OBC--PBC analysis for phase diagram}}

\begin{figure}[htbp]
    \centering
    \includegraphics[width=0.99\linewidth]{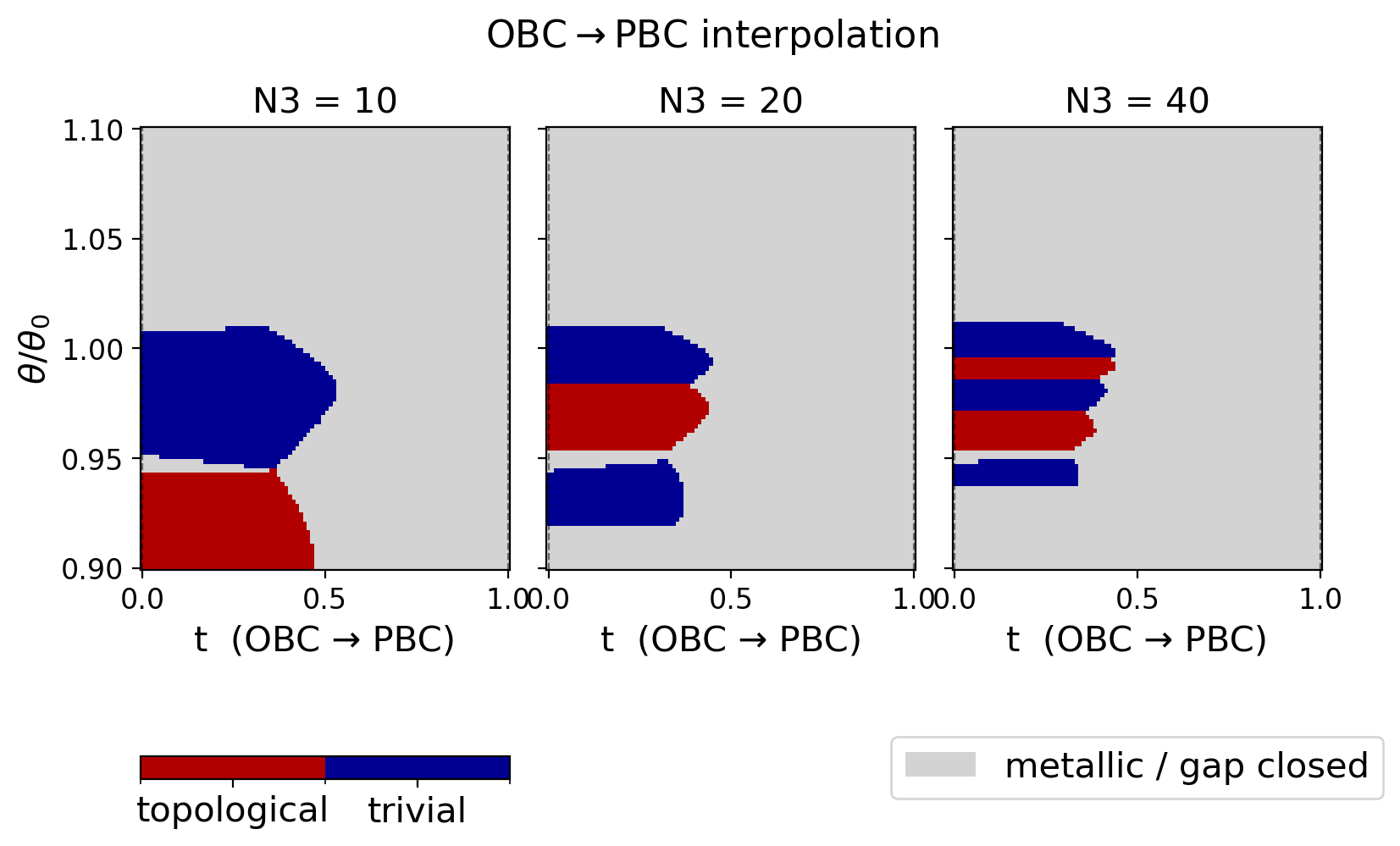}
    \caption{
        Phase diagrams in the $(\theta/\theta_0,\, t)$ plane for three slab thicknesses $N_3 = 10, 20, 40$, where $t \in [0,1]$ interpolates between open ($t=0$) and periodic ($t=1$) boundary conditions in the $N_3$ direction. Red and blue regions denote topological and trivial insulating phases, respectively, while grey regions indicate a gapless (metallic) phase with no well-defined invariant.
    }
    \label{fig:obc_pbc}
\end{figure}

{To understand the structure of the topological phase diagram in Fig.~\ref{layers:topological_phase_diagram}, we performed an additional analysis based on an interpolation between open (OBC) and periodic (PBC) boundary conditions in the confinement direction. Specifically, we introduced a parameter $t \in [0,1]$ that continuously scales the wrap-around hopping connecting the outermost layers, with $t=0$ corresponding to the physical slab geometry and $t=1$ to a fully periodic cylinder. The resulting phase diagrams in the $(\theta/\theta_0,\, t)$ plane are shown in Fig.~\ref{fig:obc_pbc} for three slab thicknesses $N_3 = 10, 20, 40$.}

{Two key features are immediately apparent. For all thicknesses, the topologically non-trivial and trivial insulating regions collapse into a gapless (metallic) phase for $t \gtrsim 0.4$, demonstrating that the open surface termination is essential for stabilizing the bulk gap. This collapse is attributed to the hybridization of surface states on opposite faces through the wrap-around hopping, which closes the gap even at moderate $t$.}

{As the thickness increases from $N_3 = 10$ to $N_3 = 40$, the single connected topological lobe visible at small thickness progressively splits into alternating stripes of topological and trivial phases. This stripe pattern is a direct fingerprint of quantum confinement: the discretization of bulk bands into subbands by the finite thickness causes each successive subband to cross the band inversion condition at a slightly different strain threshold, producing oscillatory behavior of the invariant with $N_3$. We therefore conclude that the narrowing of the topological phase window with increasing thickness is primarily driven by quantum confinement rather than interlayer coupling, with the surface termination playing an additional role in stabilizing the gapped phase.}
\medskip
\bibliography{HgTe,QW}
\end{document}